\documentclass[acmsmall,screen]{acmart}
\setcopyright{rightsretained}
\acmPrice{}
\acmJournal{PACMPL}
\acmVolume{4}
\acmNumber{OOPSLA} % CONF = POPL or ICFP or OOPSLA
\acmArticle{200}
\acmYear{2020}
\copyrightyear{2020}
\acmSubmissionID{oopsla20main-p304-p}
\acmMonth{11}
\acmDOI{10.1145/3428268}
\startPage{1}

\usepackage{booktabs}   %% For formal tables:
\usepackage{subcaption} %% For complex figures with subfigures/subcaptions
\usepackage{amsmath}
\usepackage{amsthm}
\usepackage{mathpartir}
\usepackage{listings}
\usepackage{wrapfig}
\usepackage{subcaption}
\usepackage{mdframed}
\usepackage{endnotes}

\lstdefinelanguage{imp}{
    language=C,
    morekeywords={choose, infer, observe, input, assert, skip, sample}
}

\newcommand{\code}[1]{{\ensuremath{\; \mathrm{\texttt{{#1}}} \;}}}
\newcommand{\rcode}[1]{{\ensuremath{\; \mathrm{\texttt{{#1}}}}}}
\newcommand{\lcode}[1]{{\ensuremath{\mathrm{\texttt{{#1}}} \;}}}
\newcommand{\ncode}[1]{{\ensuremath{\mathrm{\texttt{{#1}}}}}}

\newcommand{\expr}{e}
\newcommand{\Expr}{E}

\newcommand{\prop}{p}
\newcommand{\Prop}{P}

\newcommand{\modalprop}{\prop_\Box}
\newcommand{\ModalProp}{\Prop_\Box}

\newcommand{\existsprop}{\prop_\exists}
\newcommand{\ExistsProp}{\Prop_\exists}

\newcommand{\statement}{s}
\newcommand{\Statement}{S}

\newcommand{\var}{x}
\newcommand{\Var}{X}

\newcommand{\qvar}{y}

\newcommand{\bool}{b}

\newcommand{\num}{c}
\newcommand{\Num}{\mathcal{C}}

\newcommand{\true}{\mathrm{\texttt{true}}}

\newcommand{\pstate}{\sigma}
\newcommand{\PState}{\Sigma}

\newcommand{\optpstate}{\mu}

\newcommand{\bstate}{\beta}

\newcommand{\error}{\bot}

\newcommand{\config}{C}

\newcommand{\obslst}{o}
\newcommand{\ObsLst}{O}

\newcommand{\tuple}[1]{{\langle {#1} \rangle}}

\newcommand{\htriplepc}[4]{{#1} \vdash \{{#2}\} \; {#3} \; \{{#4}\}}

\newcommand{\pc}{PC}
\newcommand{\ndet}{N}
\newcommand{\determ}{D}

\newcommand{\langname}{BLIMP}

\newcommand{\bnf}{\mathrm{\texttt{::=}}}

\newcommand{\concat}{%
    \mathbin{{+}\mspace{-5mu}{+}}%
}

\newtheorem{property}{Property}

\definecolor{morange}{RGB}{255,100,0}
\definecolor{mblue}{RGB}{86,180,233}
\definecolor{mgreen}{RGB}{0,158,115}
\definecolor{myellow}{RGB}{225,225,0}
\definecolor{mred}{RGB}{255,255,0}

\mdfdefinestyle{mdfhl}{hidealllines=true,innerleftmargin=0,innerrightmargin=0,innertopmargin=0,innerbottommargin=0,backgroundcolor=yellow!20}
\begin{document}

%% Title information
%\title[Short Title]{Full Title}         %% [Short Title] is optional;
%                                        %% when present, will be used in
%                                        %% header instead of Full Title.
%\titlenote{with title note}             %% \titlenote is optional;
%                                        %% can be repeated if necessary;
%                                        %% contents suppressed with 'anonymous'
%\subtitle{Subtitle}                     %% \subtitle is optional
%\subtitlenote{with subtitle note}       %% \subtitlenote is optional;
%                                        %% can be repeated if necessary;
%                                        %% contents suppressed with 'anonymous'

\title{Programming and Reasoning with Partial Observability}

%% Author information
%% Contents and number of authors suppressed with 'anonymous'.
%% Each author should be introduced by \author, followed by
%% \authornote (optional), \orcid (optional), \affiliation, and
%% \email.
%% An author may have multiple affiliations and/or emails; repeat the
%% appropriate command.
%% Many elements are not rendered, but should be provided for metadata
%% extraction tools.

%% Author with single affiliation.
\author{Eric Atkinson}
%\authornote{with author1 note}          %% \authornote is optional;
                                        %% can be repeated if necessary
%\orcid{nnnn-nnnn-nnnn-nnnn}             %% \orcid is optional
\affiliation{
  %\position{Position1}
  %\department{Department1}              %% \department is recommended
  \institution{Massachusetts Institute of Technology}            %% \institution is required
  %\streetaddress{Street1 Address1}
  %\city{City1}
  %\state{State1}
  %\postcode{Post-Code1}
  \country{USA}                    %% \country is recommended
}
\email{eatkinson@csail.mit.edu}          %% \email is recommended

%% Author with two affiliations and emails.
\author{Michael Carbin}
%\authornote{with author2 note}          %% \authornote is optional;
                                        %% can be repeated if necessary
%\orcid{nnnn-nnnn-nnnn-nnnn}             %% \orcid is optional
\affiliation{
  %\position{Position2a}
  %\department{Department2a}             %% \department is recommended
  \institution{Massachusetts Institute of Technology}           %% \institution is required
  %\streetaddress{Street2a Address2a}
  %\city{City2a}
  %\state{State2a}
  %\postcode{Post-Code2a}
  \country{USA}                   %% \country is recommended
}
\email{mcarbin@csail.mit.edu}         %% \email is recommended
%\affiliation{
%  \position{Position2b}
%  \department{Department2b}             %% \department is recommended
%  \institution{Institution2b}           %% \institution is required
%  \streetaddress{Street3b Address2b}
%  \city{City2b}
%  \state{State2b}
%  \postcode{Post-Code2b}
%  \country{Country2b}                   %% \country is recommended
%}
%\email{first2.last2@inst2b.org}         %% \email is recommended

%\authorsaddresses{}

%% Abstract
%% Note: \begin{abstract}...\end{abstract} environment must come
%% before \maketitle command
\begin{abstract}
    Computer programs are increasingly being deployed in partially-observable environments. 
A partially observable environment is an environment whose state is not completely visible to the program, but from which the program receives partial observations.
Developers typically deal with partial observability by writing a state estimator that, given observations, attempts to deduce the hidden state of the environment.
In safety-critical domains, to formally verify safety properties developers may write an environment model. The model captures the relationship between observations and hidden states and is used to prove the software correct.

In this paper, we present a new methodology for writing and verifying programs in partially observable environments.
We present {\em belief programming}, a programming methodology where developers write an environment model that the program runtime automatically uses to perform state estimation. 
A belief program dynamically updates and queries a belief state that captures the possible states the environment could be in. 
To enable verification, we present {\em Epistemic Hoare Logic} that reasons about the possible belief states of a belief program the same way that classical Hoare logic reasons about the possible states of a program. 
We develop these concepts by defining a semantics and a program logic for a simple core language called \langname{}.
In a case study, we show how belief programming could be used to write and verify a controller for the Mars Polar Lander in \langname{}.
We present an implementation of BLIMP called CBLIMP and evaluate it to determine the feasibility of belief programming.

\end{abstract}

\ccsdesc[500]{Software and its engineering~General programming languages}
\ccsdesc[300]{Software and its engineering~Semantics}
\ccsdesc[300]{Theory of computation~Operational semantics}
\ccsdesc[500]{Theory of computation~Pre- and post-conditions}
\ccsdesc[500]{Theory of computation~Invariants}
\ccsdesc[300]{Theory of computation~Modal and temporal logics}
\ccsdesc[500]{Theory of computation~Hoare logic}
\ccsdesc[500]{Theory of computation~Program reasoning}
\ccsdesc[500]{Computing methodologies~Reasoning about belief and knowledge}
%% Keywords
%% comma separated list

\keywords{programming languages, logic, partial observability, uncertainty}

%% \maketitle
%% Note: \maketitle command must come after title commands, author
%% commands, abstract environment, Computing Classification System
%% environment and commands, and keywords command.
\maketitle

\section{Introduction}

Computer systems are increasingly deployed in {\em partially observable} environments in which the system cannot exactly determine the environment's true state~\cite{aima, pomdp}.
For example, the software that controls an uncrewed aerial vehicle (UAV) cannot exactly determine the vehicle's true altitude above the ground.
Instead, the vehicle's software receives a measurement from a GPS altimeter that estimates the vehicle's altitude. 

This measurement or {\em observation} reveals only partial information about the environment's true state, such as that the UAV's true altitude is within 25 feet of the reported measurement. The primary challenge for a system deployed in such an environment is therefore that it must leverage the partial information provided by an observation to meet its goals, which, in contrast, are typically expressed in terms of the environment's true state.

Consider, for example, a UAV tasked with avoiding a collision with the ground. 
Its controller software will include a {\em state estimator} that must infer if it is possible that the vehicle may soon have an altitude of 0 given only estimated altitude measurements.
If the vehicle is indeed at risk, then the controller must take action to ensure the vehicle's altitude stays strictly positive. 
However, the discrepancy between the measurements and the vehicle's true altitude introduces the risk that the state estimator's inferences may indicate a strictly positive altitude when the true altitude is in fact~0.
Moreover, the controller's reasoning must soundly work with the state estimator's inferences to intervene whenever the true altitude is dangerously near 0.

In safety-critical domains that desire formal guarantees, such as robotics and vehicle navigation, the best-available approach to formally verify the system is to
1) formally specify an {\em environment model}: the specific relationship between an observation and the system's true state;
2) implement the state estimator and verify the correctness of its inferences in relation to the environment model; and
3) implement the remainder of the controller and verify that the composition of the environment model and controller meets the system's requirements.

\subsection{Belief Programming}

In this paper, we present {\em belief programming}, a programming methodology in which the developer writes a program for the controller that includes a specification of the environment model. From that specification, the program runtime automatically provides a state estimator, eliminating the need to manually implement the state estimator and verify its behavior against the environment~model. 

To instantiate the concepts of this programming methodology, we present a new language, BeLief IMP (\langname{}), a variant of the pedagogical language IMP~\cite{winskel}. \langname{} provides first-class abstractions for environment modeling, for observations, and to interface with the automatically generated state estimator.

\paragraph{Environment Model} The belief programming methodology extends IMP with a nondeterministic choice statement, $\ncode{x = choose(}\prop\ncode{)}$, that nondeterministically updates the program variable \texttt{x} to a value that satisfies the predicate $\prop$. 
Unlike a traditional {\tt choose} statement~\cite{back}, the value of \texttt{x} is not immediately observable to the controller.
Instead the semantics of \texttt{x} is the set of all possible values that satisfy $\prop$. 
The programming methodology permits such nondeterministic values to be composed with additional computation to produce a jointly nondeterministic and unobserved 
set of program variables whose potential values correspond to the partially observable values of the system's physical state.

\paragraph{Observations} 
To reveal the true value of an unobserved program variable, the controller must explicitly perform an observation.
Belief programming extends IMP with an observation statement, \texttt{observe y}, that makes the value of an unobserved variable \texttt{y} visible to the program. If, for example, \texttt{y} is a measurement that is derived from another unobservable value \texttt{x}, such as an altitude measurement derived from the UAV's true altitude, then the true value of \texttt{y} reveals partial information about the true value of \texttt{x}.

\paragraph{State Estimation} 
The belief programming runtime system dynamically maintains a {\em belief state} that captures the set of all possible values of all unobserved variables.
The belief programming methodology also extends IMP with an inference statement, $\lcode{infer} \modalprop$, that computes a boolean inference over the program's belief state
and enables the controller to guide its actions given the validity of a proposition. 
For example, if $\Diamond \ncode{(altitude < 1)}$ is true, which states that is possible that the UAV's altitude is less than {\tt 1} foot, then the controller can intervene to avoid a collision.

Together our abstractions for environment modeling, observations, and state estimation enable belief programming to provide runtime capabilities for state estimation that eliminate the need to implement and verify the state estimator itself.

\subsection{Epistemic Hoare Logic}

While the belief programming methodology automates the construction of a sound state estimator, a developer must still verify that the contoller's actions and state estimator soundly work together to meet the system's requirements.  To address this problem, we present the {\em Epistemic Hoare Logic} (EHL), a variant of Hoare Logic that supports modal propositions in its assertion logic that model a belief program's dynamically tracked belief state. 

EHL includes the modal propositions $\Diamond \prop$, ``it is possible that $\prop$ is true", and $\Box \prop$, ``it is always the case that $\prop$ is true", that quantify $\prop$ over the set of all possible values of the program's variables as captured by its belief state. 
These propositions, along with EHL's inference rules, enable a developer to represent the state estimator's inferences as propositions in the logic --
e.g.\ $\Diamond \ncode{(altitude < 1)}$ meaning ``it is possible that the true altitude is less than 1 foot" -- and also specify and verify the system's requirements -- 
e.g.\ $\Box \ncode{(altitude >= 1)}$ meaning ``it is always the case that the true altitude is at least 1 foot."

\subsection{Contributions}

In this paper, we present the following contributions:

\begin{itemize}
    \item {\it Belief Programming.} We introduce belief programming, a programming methodology that makes it possible for the program runtime to automatically provide a state estimator given an environment model specification. Specifically, the program runtime tracks the program's belief state: all possible values of the program's unobservable state.
    \item {\it Language.} We present the syntax and semantics of Belief IMP (\langname{}), a language designed for belief programming. We establish basic properties of \langname{} semantics that should be true of any belief programming language. Namely, we show the state estimator that a \langname{} program provides soundly and precisely captures the environment's true state.
    \item {\it Epistemic Hoare Logic.} We present Epistemic Hoare Logic for verifying properties of \langname{} programs. We show that our logic is sound with respect to \langname{}'s semantics.
    \item {\it Case Study.} We present a case study showing how belief programming can be used to develop a verified implementation of the Mars Polar Lander's flight control software. 
        The Mars Polar Lander is a lost space probe, hypothesized to have crashed into the surface of Mars during descent due to a control software error~\citep{mpl}. 
        We present a controller implemented with belief programming, and formally prove using EHL that it does not have the error that caused the MPL crash.
    \item {\it Implementation.} We evaluate the feasibility of belief programming by presenting an implementation of \langname{} in C called CBLIMP.
        Our results show that belief programming is feasible for problems in robotics and vehicle navigation domains.
\end{itemize}

The dual contributions of belief programming and Epistemic Hoare Logic enable developers to more easily program in partially observable environments where correctness is paramount.
Such developers must currently hand-write an environment model and a state estimator, and belief programming enables them to omit the state estimator.
Epistemic Hoare Logic enables developers to reason about the correctness of the resulting belief program, just as Hoare logic allows them to reason about the correctness classical hand-written control software.

\clearpage

\section{Example}
\label{sec:example}
\lstset{xleftmargin=2em,numbers=left}
\begin{wrapfigure}[16]{r}{0.65\textwidth}
\vspace{-2.5mm}
\begin{lstlisting}
cmd = 0
t_max = 10000; t = 0;
while (t < t_max)
{
    input obs;$\label{code:readobs}$
    // Controller loop start
    if (obs < 475) { cmd = 50 } $\label{code:climb}$
    else if (obs > 525) { cmd = -50 } $\label{code:descend}$
    else { cmd = 0 }
    // Controller loop end
    t = t + 1
}
\end{lstlisting}
\vspace{-5mm}
\caption{An altitude controller for a UAV.}
\label{fig:dronecode}
\end{wrapfigure}

In this section we show how a developer uses belief programming to implement a controller for a UAV. 
The controller's objective is to maintain the UAV's altitude at 500 feet above the ground. 
While the UAV can precisely control its altitude, it has to contend with measurement error from its altitude sensing equipment and wind gusts that can blow it off course.

The listing in Figure~\ref{fig:dronecode} shows how a developer can write such a controller in a traditional programming language.
At every time step, up to a maximum of $\ncode{10000}$ steps, the controller receives an altitude observation \lstinline{obs} (Line~\ref{code:readobs}).
We assume the observation comes from a sensor which has some inherent measurement error, so that the value stored in \lstinline{obs} is not precisely the UAV's true altitude.
If the observation is sufficiently low, the controller issues a command to climb by 50 feet (Line~\ref{code:climb}).
Conversely, if the observation is sufficiently high, the controller issues a command to descend (Line~\ref{code:descend}).
The conditions on Lines \ref{code:climb} and \ref{code:descend} form a coarse-grained state estimator that determines if the UAV is too high, too low, or at an acceptable altitude.
We assume the command is stored in \lstinline{cmd}, and that there is an external process that reads \lstinline{cmd} and modifies the UAV's altitude by exactly \lstinline{cmd}.

\begin{wrapfigure}[21]{l}{0.55\textwidth}
\begin{lstlisting}
alt = 500; cmd = 0;
t_max = 10000; t = 0;
while (t < t_max)
{ 450 <= alt && alt <= 550 }$\label{code:loopinv}$
{
    alt = choose(alt - 25 <= . && $\label{code:choosealt}$
        . <= alt + 25);
    obs = choose(alt - 25 <= . && $\label{code:chooseobs}$
        . <= alt + 25);
    // Controller loop start
    ...$\label{code:droneinline}$
    // Controller loop end
    alt = alt + cmd$\label{code:updatealt}$
    t = t + 1;
}
\end{lstlisting}
\caption{Environment model for the UAV altitude controller.}
\label{fig:dronemodel}
\end{wrapfigure}

The developer needs to ensure the UAV maintains a consistent altitude for safety reasons.
We will assume that the developer wants to provide this assurance using formal verification, which requires an environment model of how the true and observed altitude are related.
The developer can write such a model as a program that specifies the set of environments the UAV may be in, including the set of values the UAV's true and observed altitude may take on.

The listing in Figure~\ref{fig:dronemodel} shows how a developer can write such a model.
The model is composed with the controller by inlining the annotated lines in Figure~\ref{fig:dronecode} into Line~\ref{code:droneinline} of Figure~\ref{fig:dronemodel}.
Note that this replaces the input \lstinline[breaklines=true]{obs} in Figure~\ref{fig:dronecode} with the value of \lstinline{obs} specified by the model, and we have added a \emph{loop invariant} on Line~\ref{code:loopinv}.

The modeling language includes a nondeterministic assignment operator \lstinline{choose}, which takes a predicate and nondeterministically chooses a value that satisfies that predicate.
The placeholder \lstinline{.} stands for the new value of the variable, so that \lstinline[breaklines=true]{x = choose(x - 1 <= . && . <= x + 1)} picks a new value for \lstinline{x} that is within a distance of \lstinline{1} of its previous value.

At every time step, the model chooses the current altitude \lstinline{alt} from a value within a distance of \lstinline{25} of the previous altitude (Line~\ref{code:choosealt}). 
This models a wind gust causing a change of up to 25 feet of altitude per step.
The model then chooses the observation \lstinline{obs} from within a distance of \lstinline{25} of the true altitude (Line~\ref{code:chooseobs}). 
This models the altitude instrumentation as having a measurement error of up to 25 feet.
After the controller runs, the model alters the altitude by adding to it the resulting command \lstinline{cmd} (Line~\ref{code:updatealt}).

The condition the developer needs to ensure is given by the loop invariant on Line~\ref{code:loopinv}: \lstinline[breaklines=true]{450 <= alt && alt <= 550}.
This means that the UAV maintains its target altitude of 500 feet within an error margin of 50 feet.
The developer can prove that the composition of the environment model and the controller satisfy this condition using classical verification techniques.

\subsection{Belief Programming}

\begin{figure}
\begin{lstlisting}
alt = 500; cmd = 0; t_max = 10000; t = 0;
while (t < t_max)
{ $\Box$(450 <= alt && alt <= 550) }$\label{code:bploopinv}$
{
    alt = choose(alt - 25 <= . && . <= alt + 25);$\label{code:bpchoosei}$
    obs = choose(alt - 25 <= . && . <= alt + 25);$\label{code:bpchooseii}$
    observe obs;$\label{code:bpobserve}$
    
    infer $\Diamond$(alt < 450) { cmd = 50 } $\label{code:inferstart}$
    else infer $\Diamond$(alt > 550) { cmd = -50 } 
    else { cmd = 0 };$\label{code:inferend}$
    
    alt = alt + cmd;$\label{code:altasgn}$
    t = t + 1$\label{code:bodyend}$
}
\end{lstlisting}
\caption{Implementation of the UAV controller using belief programming in \langname{}.}
\label{fig:dronebp}
\end{figure}

We will now explain, alternatively, how the developer implements this program using belief programming.
The listing in Figure~\ref{fig:dronebp} shows the code to implement the controller in our belief programming language \langname{}.
As this program executes, it maintains a {\em belief state} of the set of possible environments it could be in.
Instead of the conditions over concrete observations in Figure~\ref{fig:dronecode}, the code in Figure~\ref{fig:dronebp} uses conditions over belief states to determine control actions.

To explain how belief programming operators work, we will walk through the execution of the first iteration of the while loop in Figure~\ref{fig:dronebp}.
At the start of the loop, the belief state contains a single environment that has $\ncode{alt} = 500$, $\ncode{cmd} = 0$, $\ncode{t\_max} = 10000$, and $\ncode{t} = 0$.

\paragraph{Choose Statements} The \lstinline{choose} statements on Lines \ref{code:bpchoosei} and \ref{code:bpchooseii} expand the belief state to include all possible environments the nondeterminism could generate.
After the first assignment to \lstinline{alt} on Line~\ref{code:bpchoosei}, the belief state contains all environments such that $\ncode{alt} \in [475, 525]$.
After the assignment to \lstinline{obs} on Line~\ref{code:bpchooseii}, the belief state contains all environments such that $\ncode{obs} \in [450, 550]$, with the additional constraint that the distance between \lstinline{alt} and \lstinline{obs} is less than or equal to $25$.
This means that, for example, the belief state does not contain the environment where $\ncode{alt} = 500$ and $\ncode{obs} = 550$.

\paragraph{Observe Statements} The \lstinline{observe} statement on Line~\ref{code:bpobserve} implicitly receives an input that is the observed altitude \lstinline{obs}.
It updates the belief state to contain only environments that are consistent with that observed altitude.
For example, if the program receives the value $525$, then \lstinline{observe} modifies the belief state
to only contain environments where $\ncode{obs} = 525$ and, correspondingly, where $\ncode{alt} \in [500, 525]$.

\paragraph{Infer Statements}
The \lstinline{infer} statements on Lines~\ref{code:inferstart}-\ref{code:inferend} branch based on belief state conditions. 
Conditions use the $\Box$ and $\Diamond$ {\em modal operators} to quantify over environments in the belief state, with $\Box$ meaning ``for all environments in the belief state'' and $\Diamond$ meaning ``there exists an environment in the belief state''.
The condition \lstinline{$\Diamond$(alt < 450)} means that there is an environment in the belief state such that \lstinline{alt} is smaller than $450$. 
Similarly \lstinline{$\Diamond$(alt > 550)} means that there is an environment such that \lstinline{alt} is larger than $550$.
Compared to the state estimator on Lines \ref{code:climb} and \ref{code:descend} of Figure~\ref{fig:dronecode}, the infer statements provide more intuition for why the controller was constructed this way. 
If one of these conditions is true, then it is possible for the true environment to be outside of the desired range of $500 \pm 50$ feet, and immediate action is needed to correct the situation.
Since, assuming the example observation of $525$, our belief state contains only the environments where $\ncode{alt} \in [500, 525]$, neither of these conditions is true.
Thus, the \lstinline{cmd = 0} branch of the infer statements is executed. 
This constrains every environment in the belief state to include \lstinline{cmd = 0}.
Note that under the assumption that the observation localizes $\ncode{alt}$ to within 25 feet, at most one condition of the infer statements will be true in any given belief state.

\paragraph{Assignments} The final line of the loop, Line~\ref{code:altasgn}, updates \lstinline{alt} to be its previous value plus the value of \lstinline{cmd} in every environment in the belief state. 
Because the belief state contains the environments such that $\ncode{cmd} = 0$ and $\ncode{alt} \in [500, 525]$, after the update the belief state contains all environments such that $\ncode{alt} \in [500, 525]$.
The loop invariant on Line~\ref{code:bploopinv} states that any environment in the belief state must have a value for $\ncode{alt}$ in the range $[450, 550]$.
Because our belief state constrains $\ncode{alt}$ to the smaller range $[500, 525]$, our belief state satisfies the invariant.

\subsection{Reasoning with Epistemic Hoare Logic}

The previous section describes a single execution of the belief program given concrete observations, but in general developers need to reason about all potential executions given any observations that are allowed under the environment model.
In this section, we show how a developer can use Epistemic Hoare Logic to reason about the potential executions of the belief program in Figure~\ref{fig:dronebp}.
We first present a small example that showcases how Epistemic Hoare Logic can be used to reason about a single statement in the program.
We then explain at a high level how similar reasoning can be used to verify the program maintains its altitude within limits, with the details in Appendix~\ref{sec:exampleverification}.

\subsubsection{Small Example}
\newcommand{\stmtref}{\statement^*}

In this section, we consider proving a simple property about the statement on Line~\ref{code:bpchoosei} of Figure~\ref{fig:dronebp} that specifies altitude updates.
We will assume, in accordance with the loop invariant, that the altitude immediately before executing this statement is larger than 450.
We will then show that after executing this statement, the altitude is larger than 425.
This property is expressed by the following judgment in Epistemic Hoare Logic:
\[
    \htriplepc{\determ}{\Box\ncode{(450 <= alt)}}
        {\ncode{alt = choose(alt - 25 <= . \&\& . <= alt + 25)}}
        {\Box\ncode{(425 <= alt)}}
\]
The context $\determ$ indicates that the judgement applies when the statement is executed under deterministic control flow.
The pre-condition, $\Box\ncode{(450 <= alt)}$, states that in any environment in the belief state, the variable \texttt{alt} is at least \texttt{450}.
The post-condition, $\Box\ncode{(425 <= alt)}$, states that in any environment in the belief state, the variable \texttt{alt} is at least \texttt{425}.

\paragraph{Epistemic Hoare Logic Rules.} To prove this deduction, we first instantiate the Epistemic Hoare Logic rule for \texttt{choose} statements that we have designed.
We present the general rule in Figure~\ref{fig:logic} of Section~\ref{sec:logic}.
By default, the rule allows for nondeterministic control flow, but we can specialize it to the $\determ$ context via a subtyping rule.
Instantiating the choose rule and subtyping with our original pre-condition yields the following judgement, using the notation $\stmtref$ for the statement on Line~\ref{code:bpchoosei}:
\[
    \htriplepc{\determ}{\Box\ncode{(450 <= alt)}}
        {\stmtref}
        {\Box\ncode{(450 <= a)} \code{\&\&} 
            \Box\ncode{(a - 25 <= alt \&\& alt <= a + 25)}}
\]
The rule produces a post-condition by first renaming all instances of the assigned variable \texttt{alt} with a fresh variable \texttt{a} in the pre-condition.
It then conjuncts this with the \texttt{choose} statement's predicate under a $\Box$ with \texttt{alt} replaced with \texttt{a} and the placeholder \lstinline{.} replaced with \texttt{alt}.

\paragraph{Implications.}
To rewrite the post-condition into the desired form, we prove that another predicate is implied by the post-condition and use the rule of consequence (formalized in general in Figure~\ref{fig:logic}) to replace the post-condition with the new predicate.

First, we use the general principle that the $\Box$ operator commutes with \texttt{\&\&}.
We have formalized this principle as Theorem~\ref{thm:box-and} in Appendix~\ref{sec:proptheorems}.
This gives us the following judgment:
\[
    \htriplepc{\determ}{\Box\ncode{(450 <= alt)}}
        {\stmtref}
        {\Box\ncode{(450 <= a \&\& a - 25 <= alt \&\& alt <= a + 25)}}
\]

We then use the principle of lifting theorems about environments to theorems about belief states.
This states that if we have an implication over environments, we can wrap the premise and conclusion of this implication with $\Box$, and obtain an implication over belief states.
We have formalized this principle as Theorems \ref{thm:lifting} and \ref{thm:k} in Appendix~\ref{sec:proptheorems}.
In this example, we use the fact that if the environment satisfies \lstinline[breaklines=true]{450 <= a && a - 25 <= alt && alt <= a + 25}, then it also satisfies \lstinline[breaklines=true]{425 <= alt}.
The principle of lifting says that as a result,
if the belief state satisfies \lstinline[breaklines=true]{$\Box$(450 <= a && a - 25 <= alt && alt <= a + 25)}, then it also satisfies \lstinline[breaklines=true]{$\Box$(425 <= alt)}.
Applying the rule of consequence gives us the original judgment we set out to prove:
\[
    \htriplepc{\determ}{\Box\ncode{(450 <= alt)}}
        {\stmtref}
        {\Box\ncode{(425 <= alt)}}
\]

\subsubsection{Verification}

The developer would like to ensure that the loop body in Figure~\ref{fig:dronebp} maintains an altitude of 500 feet within a tolerance of 50 feet.
This means that the loop body must preserve its invariant.
This corresponds to the Epistemic Hoare Logic deduction
\[
    \htriplepc{\determ}{\Box\ncode{(450 <= alt \&\& alt <= 550)}}{s}{\Box\ncode{(450 <= alt \&\& alt <= 550)}}
\]
where the pre- and post-conditions are both equal to the loop invariant on Line~\ref{code:bploopinv} of Figure~\ref{fig:dronebp}.
The program $s$ is the loop body on Lines~\ref{code:bpchoosei}-\ref{code:bodyend} of Figure~\ref{fig:dronebp}.
Thus, this deduction states that if the loop body  starts in a belief state satisfying the loop invariant, it produces a belief state that also satisfies the loop invariant.
We consider the case of deterministic control flow, as signified with the $\determ$ context.
This is appropriate because the loop condition, \texttt{t < t\_max}, is consistently either true or false across all environments in the belief state.

The high-level procedure is to assume the loop invariant as the pre-condition for the loop body. 
We then use the logic's rules to derive an appropriate post-condition for the loop body, and finally prove that the post condition implies the loop invariant using the reasoning principles in Section~\ref{sec:proptheorems}.
The details of this process are in Appendix~\ref{sec:exampleverification}.

\section{Language: Syntax and Semantics of \langname{}}
\label{sec:language}
\newcommand{\sep}{\; \mid \;}
\newcommand{\lsep}{\mid \;}
\newcommand{\mkproplang}[2]{{%
    {#2} \sep {#1} \code{\&\&} {#1} \sep {#1} \code{||} {#1} \sep %
    \ncode{!} {#1}
}}
\begin{figure}
    \begin{align*}
        \Expr & \; \bnf \; \var \sep \num \sep \Expr \code{+} \Expr %
            \sep \Expr \code{-} \Expr \sep \Expr \code{*} \Expr \sep%
            \Expr \code{/} \Expr \sep \qvar \sep \ncode{.} \\
        \Prop & \; \bnf \; \mkproplang{\Prop}{\bool \sep%
            \Expr \code{<} \Expr \sep \Expr \code{==} \Expr}\\
        \ModalProp & \; \bnf \; \mkproplang{\ModalProp}{\Box \Prop \sep \Diamond \Prop}\\
        \ExistsProp & \; \bnf \; \exists \qvar^*. \; \ModalProp 
    \end{align*}
    \caption{Syntax of expressions and propositions.}
    \label{fig:propsyntax}
\end{figure}

In this section, we present the belief programming language \langname{}. We first present \langname{}'s syntax and semantics, and then state and prove properties of the semantics.
We then present an {\em execution model}. 
While the semantics describes both the environment modeling and state estimation behavior of \langname{} programs, the execution model projects out the state estimation operations.

\subsection{Syntax}

In this section, we present the syntax of \langname{}, which gives the constructs of belief programming.

\subsubsection{Expressions and Propositions} Figure~\ref{fig:propsyntax} gives the syntax of expressions and propositions. 

\paragraph{Expressions.} We use the notation $\Expr$ to refer to expressions. 
An expression may be a variable $\var$, a numeric constant $\num$, or formed using one of the binary operators \rcode{+}, \rcode{-}, \rcode{*}, \rcode{/}, which have standard interpretations.
An expression may also contain a quantified variable $\qvar$ or the placeholder \code{.}.
Quantified variables only make sense when the variable is bound by an outer quantifier, and the placeholder value only makes sense in the context of an enclosing \texttt{choose} statement.

\paragraph{Propositions.} We use the notation $\Prop$ to refer to propositions.
Propositions may be boolean constants $\bool \in \{\ncode{true}, \ncode{false}\}$ or comparisons between expressions using the comparison operators $\ncode{<}$ and $\ncode{==}$.
Propositions may also be combined by conjunction, disjunction, and negation through the boolean operators $\ncode{\&\&}$, $\ncode{||}$, and $\ncode{!}$, respectively.

\paragraph{Modal Propositions.} We use the notation $\ModalProp$ to refer to modal propositions. 
Modal propositions are propositions that are modified using the $\Box$ and $\Diamond$ operators. 
They quantify over environments in the belief state, and are used to query the belief state for state estimation.
Modal propositions may be combined by conjunction, disjunction, and negation, using the same syntax as for non-modal propositions.
Note that in contrast to many modal logics, \langname{}'s modal operators may only be applied once; hence, propositions such as $\Box \Box \Prop$ are not in the language.
This is a design decision we have made to succinctly capture properties of the domain.
The alternative would be to allow propositions such as $\Box \Box \Prop$, and include a theorem of the form $\Box \Box \Prop \Rightarrow \Box \Prop$.

\paragraph{Existential Propositions.} We use the notation $\ExistsProp$ to refer to existentially quantified modal propositions.
An existential proposition is a modal proposition prepended with the $\exists$ symbol and a comma-separated list of quantified variables $\qvar^*$ (which could potentially be empty).
Existential propositions form the core propositional language for reasoning with Epistemic Hoare Logic, and appear in \langname{} in assertions and loop invariants.

\subsubsection{Statements} Figure~\ref{fig:statementsyntax} presents the syntax of statements. We use the notation $\Statement$ to refer to the set of statements, which is specified by the grammar in Figure~\ref{fig:statementsyntax}. A statement may be an assignment, a choose statement, an assertion, an observation, an if statement, an infer statement, a while loop, a composition of two statements, or a skip statement.

\begin{wrapfigure}[14]{l}{0.45\textwidth}
    \begin{minipage}{0.45\textwidth}
    \begin{align*}
        \Statement & \; \bnf \; \var \code{=} \Expr \\
          & \sep \var \code{= choose(} \Prop \rcode{)}\\
          & \sep \lcode{assert} \ExistsProp\\
          & \sep \lcode{observe} \var\\
          & \sep \lcode{if} \Prop \code{\{} \Statement \code{\} else \{} \Statement \code{\}}\\
          & \sep \lcode{infer} \ModalProp \code{\{} \Statement \code{\} else \{} \Statement \code{\}}\\
          & \sep \code{while} \Prop \code{\{} \ExistsProp \code{\} \{} \Statement \code{\}}\\
          & \sep \Statement \code{;} \Statement\\
          & \sep \ncode{skip}
    \end{align*}
    \end{minipage}
    \caption{Syntax of statements}
    \label{fig:statementsyntax}
\end{wrapfigure}

\paragraph{Assignment and Choose.} An assignment statement $\var \code{=} \Expr$ and a choose statement $\var \code{= choose(} \Prop \rcode{)}$ both assign to the program variable $\var$. 
The assignment statement does so using an expression while the choose statement does so using a proposition that contains the $\ncode{.}$ placeholder value.

\paragraph{Assertions.} The keyword \texttt{assert} signifies an assertion statement. 
An assertion includes an existential proposition specifies a property that must be true at the statement's program point.

\paragraph{Observation.} The keyword \texttt{observe} signifies an observation statement. An observation includes the program variable, $\var$, to be observed at the statement's program point.

\paragraph{If and Infer Statements.} The keywords \texttt{if} and \texttt{infer} signify an if statement and an infer statement, respectively. 
Both statements select branches based on a condition. 
The distinction between an if and an infer statement is that an if statement branches on an ordinary proposition, whereas an infer statement branches on a modal proposition.
If statements facilitate conditional environment models, whereas infer statements facilitate state estimation by branching on belief state queries.

\paragraph{While Loops.} The keyword \texttt{while} signifies a while loop. 
Such a loop consists of a body, a condition regarding whether to continue or not, and a loop invariant. 
The loop invariant is an existential proposition that must be true at the start and end of each loop, and may be used to express a safety property that must be true at every iteration of a time-step loop.

\paragraph{Composition and Skip.} Statements may be sequentially composed with the $\ncode{;}$ operator. 
The \texttt{skip} keyword signifies a skip statement, a null statement that performs no operations.

\subsection{Semantics}
\label{sec:semantics}

In this section, we illustrate how belief programming works by presenting the semantics of \langname{}. 
The semantics precisely specify how \langname{} manipulates belief states with the goal that the belief states always capture all possible behaviors and only capture realizable behaviors.

\subsubsection{Preliminaries}

\paragraph{Environments.} An environment $\pstate \in \PState = (\Var \rightarrow \Num)$ is a finite map from program variable names to their numeric values, where each value belongs to a finite subset $\Num$  of the integers.
We use the notation $\pstate[\var \mapsto \num]$ to mean the environment $\pstate$ with the variable $\var$ mapped to the value $\num$.
An \emph{optional environment} $\mu \in \PState \cup \{ \cdot \}$ may be either an environment $\pstate$ or a null value $\cdot$.
The use of null optional environments is described in more detail in Section~\ref{sec:statementsemantics}.

\paragraph{Belief States.} A belief state $\bstate \in \mathcal{P}(\PState)$ is an element of the powerset of $\PState$, i.e.\ it is a set of environments. 
The interpretation is that if an environment $\pstate$ is in $\bstate$, then the program believes that $\pstate$ is a possibly true environment.

\subsubsection{Expressions}

\begin{figure}
    \begin{mathpar}
        \inferrule{ }
        {\tuple{\pstate, \var} \Downarrow \pstate(\var)}

        \inferrule{ }
        {\tuple{\pstate, \num} \Downarrow \num}

        \inferrule{ \tuple{\pstate, \expr_0} \Downarrow \num_0\\\\
                    \tuple{\pstate, \expr_1} \Downarrow \num_1
        }
        {
            \tuple{\pstate, \expr_0 \code{+} \expr_1} \Downarrow \num_0 + \num_1
        }

        \inferrule{ \tuple{\pstate, \expr_0} \Downarrow \num_0\\\\
                    \tuple{\pstate, \expr_1} \Downarrow \num_1
        }
        {
            \tuple{\pstate, \expr_0 \code{-} \expr_1} \Downarrow \num_0 - \num_1
        }

        \inferrule{ \tuple{\pstate, \expr_0} \Downarrow \num_0\\
                    \tuple{\pstate, \expr_1} \Downarrow \num_1
        }
        {
            \tuple{\pstate, \expr_0 \code{*} \expr_1} \Downarrow \num_0 * \num_1
        }

        \inferrule{ \tuple{\pstate, \expr_0} \Downarrow \num_0\\
                    \tuple{\pstate, \expr_1} \Downarrow \num_1
        }
        {
            \tuple{\pstate, \expr_0 \code{/} \expr_1} \Downarrow \num_0 / \num_1
        }
    \end{mathpar}

\caption{Semantics of expressions. We use the notation $\tuple{\pstate, \expr} \Downarrow \num$ to mean that the expression $\expr$ evaluated under the environment $\pstate$ yields the value $\num$.}
\label{fig:exprsemantics}
\end{figure}

Figure~\ref{fig:exprsemantics} presents the semantics of expressions.
Our approach is a big-step operational semantics that states what value an expression computes when evaluated under a given environment. 
We use the notation $\tuple{\pstate, \expr} \Downarrow \num$ to mean that the expression $\expr$ evaluated under the environment $\pstate$ yields the value $\num$.
The meaning of a variable $\var$ is the value of the variable in the input environment $\pstate$.
The meaning of a constant $\num$ is the value of the constant.
The arithmetic operators $\ncode{+}$, $\ncode{-}$, $\ncode{*}$, and $\ncode{/}$ have their standard interpretation.
We assume that $\ncode{/}$ denotes integer division.

\subsubsection{Propositions}

\begin{figure}
    \begin{align*}
        \pstate \vDash \bool & \iff \bool = \true\\
        \pstate \vDash \prop_1 \code{\&\&} \prop_2 & \iff%
            \pstate \vDash \prop_1 \wedge \pstate \vDash \prop_2\\
        \pstate \vDash \prop_1 \code{||} \prop_2 & \iff%
            \pstate \vDash \prop_1 \vee \pstate \vDash \prop_2\\
        \pstate \vDash \rcode{!} \prop & \iff \pstate \not \vDash \prop\\
        \pstate \vDash \expr_1 \code{<} \expr_2 & \iff%
            \mathrm{\textnormal{there exist}} \; \num_1, \num_2 \; \mathrm{s.t.} \;%
                \tuple{\pstate, \expr_1} \Downarrow \num_1 \wedge%
                \tuple{\pstate, \expr_2} \Downarrow \num_2 \wedge%
                \num_1 < \num_2\\
        \pstate \vDash \expr_1 \code{==} \expr_2 & \iff%
            \mathrm{\textnormal{there exist}} \; \num_1, \num_2. \; \mathrm{s.t.} \;%
                \tuple{\pstate, \expr_1} \Downarrow \num_1 \wedge%
                \tuple{\pstate, \expr_2} \Downarrow \num_2 \wedge%
                \num_1 = \num_2       
    \end{align*}
    \caption{Semantics of propositions. We use the notation $\pstate \vDash \prop$ to mean $\pstate$, which must be an environment, satisfies $\prop$, a proposition.}
    \label{fig:propsemantics}
\end{figure}

Figure~\ref{fig:propsemantics} presents the semantics of propositions.
The meaning of a proposition is whether or not, for a given environment, the environment satisfies the proposition (i.e.\ the proposition is true in the environment).
We use the notation $\pstate \vDash \prop$ to mean that the environment $\pstate$ satisfies the proposition $\prop$.
The meaning of an equality $\ncode{==}$ or size $\ncode{<}$ comparison is that an environment satisfies the comparison iff evaluating the expressions yields numbers that satisfy the comparison.
Note that the existential quantifiers in the definition are trivial because the expression semantics is a total function of the environment.
The meaning of each of the operators $\ncode{\&\&}$, $\ncode{||}$, and $\ncode{!}$ is its standard interpretation in propositional logic.

\subsubsection{Modal and Existential Propositions}
\begin{figure}[ht]
    \begin{subfigure}{0.45\textwidth}
    \begin{align*}
        \bstate \vDash \Box \prop & \iff%
            \mathrm{\textnormal{for every}} \; \pstate \in \bstate, \; \pstate \vDash \prop\\
        \bstate \vDash \Diamond \prop & \iff%
            \mathrm{\textnormal{there is some}} \; \pstate \in \bstate \; \mathrm{s.t.} \;  \pstate \vDash \prop
    \end{align*}
    \end{subfigure}
    \begin{subfigure}{0.45\textwidth}
    \begin{align*}
        \bstate \vDash \exists . \; \modalprop & \iff%
            \bstate \vDash \modalprop\\
        \bstate \vDash \exists \qvar_0, \hat{\qvar}'. \; \modalprop & \iff%
            \mathrm{\textnormal{there is some}} \; \num \; \mathrm{s.t.} \;%
                \bstate \vDash \exists \hat{\qvar}'. \; \modalprop[\num/\qvar_0]
    \end{align*}
    \end{subfigure}
    \caption{Semantics of modal and existential propositions. We use the notation $\bstate \vDash \modalprop$ and $\bstate \vDash \existsprop$ to mean that the belief state $\bstate$ satisfies $\modalprop$ or $\existsprop$.}
    \label{fig:modalpropsemantics}
\end{figure}

Figure~\ref{fig:modalpropsemantics} presents the semantics of modal and existential propositions.
The meaning of a modal or existential proposition is whether or not a given belief state satisfies the proposition (i.e.\ the proposition is true in the belief state).
We use the notation $\bstate \vDash \modalprop$ and $\bstate \vDash \existsprop$ to mean that the belief state $\bstate$ satisfies $\modalprop$ or $\existsprop$.
The meaning of $\Box$ is to universally quantify over all environments in the belief state, and the meaning of $\Diamond$ is to existentially quantify over environments in the belief state.
The meaning of the operators $\ncode{\&\&}$, $\ncode{||}$, and $\ncode{!}$ (elided in Figure~\ref{fig:modalpropsemantics}) is their standard propositional logic interpretations, and is the same as Figure~\ref{fig:propsemantics} with $\pstate$ replaced with $\bstate$.
The meaning of $\exists$ is its standard meaning in first-order logic. 
We use notation $\modalprop[\num/\qvar_0]$ to mean the proposition $\modalprop$ with $c$ substituted for $\qvar_0$, and the notation $\exists. \; \modalprop$ to mean an existential proposition with an empty set of quantified variables.

\subsubsection{Statements}
\label{sec:statementsemantics}

\begin{figure}
    \begin{mathpar}
       \inferrule
            {
                \bstate' = \{\pstate_\bstate[\var \mapsto \num_\bstate] \; | \;
                    \pstate_\bstate \in \bstate \wedge
                    \tuple{\pstate_\bstate, \expr} \Downarrow \num_\bstate
                \}
            }
            {
                \tuple{\bstate, \cdot, \var \code{=} \expr} \Downarrow 
                    \tuple{\bstate', \cdot \; | \; \ncode{nil}}
            }

       \inferrule
            {
                \tuple{\bstate, \cdot, \var \code{=} \expr} \Downarrow 
                    \tuple{\bstate', \cdot \; | \; \ncode{nil}}\\
                \tuple{\pstate, \expr} \Downarrow \num
            }
            {
                \tuple{\bstate, \pstate, \var \code{=} \expr} \Downarrow 
                    \tuple{\bstate', \pstate[\var \mapsto \num] \; | \; \ncode{nil}}
            }

        \inferrule
            {
                \bstate' = \{ \pstate_\bstate[\var \mapsto \num_\bstate] \; | \;
                    \pstate_\bstate \in \bstate \wedge
                    \pstate_\bstate \vDash \prop[\num_\bstate/\ncode{.}]
                \}
            }
            {
                \tuple{\bstate, \cdot, \var \rcode{= choose(} \prop \ncode{)}}
                    \Downarrow \tuple{\bstate', \cdot %
                    \; | \; \ncode{nil}}
            }

        \inferrule
            {
                \tuple{\bstate, \cdot, \var \rcode{= choose(} \prop \ncode{)}}
                    \Downarrow \tuple{\bstate', \cdot %
                    \; | \; \ncode{nil}}\\\\
                \pstate \vDash \prop[\num/\ncode{.}]
            }
            {
                \tuple{\bstate, \pstate, \var \rcode{= choose(} \prop \ncode{)}}
                    \Downarrow \tuple{\bstate', \pstate[\var \mapsto \num] %
                    \; | \; \ncode{nil}}
            }

        \inferrule
            {
                \bstate \vDash \modalprop
            }
            {
                \tuple{\bstate, \optpstate, \lcode{assert} \modalprop} \Downarrow
                    \tuple{\bstate, \optpstate \; | \; \ncode{nil}}
            }

        \inferrule
            {
                \bstate \not \vDash \modalprop
            }
            {
                \tuple{\bstate, \optpstate, \lcode{assert} \modalprop} \Downarrow
                    \tuple{\error \; | \; \ncode{nil}}
            }

        \inferrule
            {
                \bstate' = \{ \pstate_\bstate \; | \; \pstate_\bstate \in \bstate
                    \wedge \pstate_\bstate(\var) = \pstate(\var)\}
            }
            {
                \tuple{\bstate, \pstate, \lcode{observe} \var} \Downarrow
                    \tuple{\bstate', \pstate \; | \; \var \code{:} \pstate(\var)}
            }

        \inferrule
            { }
            {
                \tuple{\bstate, \cdot, \lcode{observe} \var} \Downarrow
                    \tuple{\error \; | \; \ncode{nil}}
            }

        \inferrule
            {
                \bstate \vDash \Diamond p \code{\&\&} \Diamond \ncode{(!}p\ncode{)}\\\\
                \bstate_T = \{ \pstate_\bstate \; | \; 
                    \pstate_\bstate \in \bstate \wedge 
                    \pstate_\bstate \vDash \prop \}\\
                \bstate_F = \{ \pstate_\bstate \; | \; 
                    \pstate_\bstate \in \bstate \wedge 
                    \pstate_\bstate \not \vDash \prop \}\\\\
                \optpstate_T = \pstate \; \mathrm{if} \; \optpstate = \pstate \wedge 
                    \pstate \vDash \prop \; \mathrm{else} \; \cdot \\
                \optpstate_F = \pstate \; \mathrm{if} \; \optpstate = \pstate \wedge 
                    \pstate \not \vDash \prop \; \mathrm{else} \; \cdot \\\\
                \tuple{\bstate_T, \optpstate_T, \statement_1} \Downarrow 
                    \tuple{\bstate_T', \optpstate_T' \; | \; \ncode{nil}}\\
                \tuple{\bstate_F, \optpstate_F, \statement_2} \Downarrow 
                    \tuple{\bstate_F', \optpstate_F' \; | \; \ncode{nil}}\\\\
                \optpstate' = {\begin{cases}
                    \optpstate_T' & \optpstate = \pstate \wedge \pstate \vDash \prop\\
                    \optpstate_F' & \optpstate = \pstate \wedge \pstate \not \vDash \prop\\
                    \cdot & \mathrm{else}
                \end{cases}}
%                \optpstate_T = \optpstate_T' \; \mathrm{if} \; 
%                    \optpstate = \pstate \wedge \pstate \vDash \prop 
%                    \; \mathrm{else} \; \optpstate_F' \\
            }
            {
                \tuple{\bstate, \optpstate, \lcode{if} \prop \code{\{} \statement_1
                    \code{\} else \{} \statement_2 \rcode{\}}} \Downarrow
                    \tuple{\bstate_T' \cup \bstate_F', 
                        \optpstate' \; | \; \ncode{nil}}
            }

        \inferrule
            {
                \bstate \vDash \Diamond p \code{\&\&} \Diamond \ncode{(!}p\ncode{)}\\\\
                \bstate_T = \{ \pstate_\bstate \; | \; 
                    \pstate_\bstate \in \bstate \wedge 
                    \pstate_\bstate \vDash \prop \}\\
                \bstate_F = \{ \pstate_\bstate \; | \; 
                    \pstate_\bstate \in \bstate \wedge 
                    \pstate_\bstate \not \vDash \prop \}\\\\
                \optpstate_T = \pstate \; \mathrm{if} \; \optpstate = \pstate \wedge 
                    \pstate \vDash \prop \; \mathrm{else} \; \cdot \\
                \optpstate_F = \cdot \; \mathrm{if} \; \optpstate = \pstate \wedge 
                    \pstate \vDash \prop \; \mathrm{else} \; \pstate \\\\
                \tuple{\bstate_T, \optpstate_T, \statement_1} \Downarrow 
                    \tuple{\bstate_T', \optpstate_T' \; | \; \obslst_T}\\
                \tuple{\bstate_F, \optpstate_F, \statement_2} \Downarrow 
                    \tuple{\bstate_F', \optpstate_F' \; | \; \obslst_F}\\\\
                \obslst_T \neq \ncode{nil} \vee \obslst_F \neq \ncode{nil}
%                \optpstate_T = \optpstate_T' \; \mathrm{if} \; 
%                    \optpstate = \pstate \wedge \pstate \vDash \prop 
%                    \; \mathrm{else} \; \optpstate_F' \\
            }
            {
                \tuple{\bstate, \optpstate, \lcode{if} \prop \code{\{} \statement_1
                    \code{\} else \{} \statement_2 \rcode{\}}} \Downarrow
                    \tuple{\error \; | \; \ncode{nil}}
            }

        \inferrule
            {
                \bstate \vDash \Box p\\
                \tuple{\bstate, \optpstate, \statement_1} \Downarrow
                    \tuple{\config_1 \; | \; \obslst}
            }
            {
                \tuple{\bstate, \optpstate, \lcode{if} \prop \code{\{} \statement_1
                    \code{\} else \{} \statement_2 \rcode{\}}} \Downarrow
                    \tuple{\config_1 \; | \; \obslst}
            }

        \inferrule
            {
                \bstate \vDash \Box \ncode{(!} p \ncode{)}\\
                \tuple{\bstate, \optpstate, \statement_2} \Downarrow
                    \tuple{\config_2 \; | \; \obslst}
            }
            {
                \tuple{\bstate, \optpstate, \lcode{if} \prop \code{\{} \statement_1
                    \code{\} else \{} \statement_2 \rcode{\}}} \Downarrow
                    \tuple{\config_2 \; | \; \obslst}
            }

        \inferrule
            {
                \bstate \vDash \modalprop\\
                \tuple{\bstate, \optpstate, \statement_1} \Downarrow \tuple{\config_1 \; | \; \obslst}
            }
            {
                \tuple{\bstate, \optpstate, \lcode{infer} \modalprop \code{\{}
                    \statement_1 \code{\} else \{} \statement_2 \rcode{\}}}
                    \Downarrow \tuple{\config_1 \; | \; \obslst}
            }

        \inferrule
            {
                \bstate \not \vDash \modalprop\\
                \tuple{\bstate, \optpstate, \statement_2} \Downarrow \tuple{\config_2 \; | \; \obslst}
            }
            {
                \tuple{\bstate, \optpstate, \lcode{infer} \modalprop \code{\{}
                    \statement_1 \code{\} else \{} \statement_2 \rcode{\}}}
                    \Downarrow \tuple{\config_2 \; | \; \obslst}
            }

        \inferrule
            {
                \tuple{\bstate, \optpstate, 
                \lcode{assert} \existsprop^I
                \rcode{; if(}\prop \ncode{) \{} \statement \code{;}  
                \lcode{while} \prop \code{\{} \existsprop^I
                    \code{\} \{} \statement \rcode{\}}
                \lcode{\} else \{ skip \}}
                } \Downarrow \tuple{\config \; | \; \obslst}
            }
            {
                \tuple{\bstate, \optpstate,
                    \lcode{while} \prop \code{\{} \existsprop^I
                    \code{\} \{} \statement \rcode{\}}} \Downarrow
                    \tuple{\config \; | \; \obslst}
            }

        \inferrule
            {
                \tuple{\bstate, \optpstate, \statement_1} \Downarrow 
                    \tuple{\bstate', \optpstate' \; | \; \obslst_1}\\\\
                \tuple{\bstate', \optpstate', \statement_2} \Downarrow 
                    \tuple{\bstate'', \optpstate'' \; | \; \obslst_2}
            }
            {
                \tuple{\bstate, \optpstate, \statement_1 \code{;} \statement_2}
                    \Downarrow \tuple{\bstate'',\optpstate'' \; | \; \obslst_1 \concat \obslst_2}
            }

        \inferrule{ }
            {
                \tuple{\bstate, \optpstate, \ncode{skip}} \Downarrow 
                    \tuple{\bstate, \optpstate \; | \; \ncode{nil}}
            }
    \end{mathpar}
    \caption{Semantics of statements. We use the notation $\tuple{\bstate, \optpstate, \statement} \Downarrow \tuple{\config \; | \; \obslst}$ to mean that the belief state $\bstate$ and optional true environment $\optpstate$ produce the configuration $\config$ augmented with the observation list $\obslst$.}
    \label{fig:statementsemantics}
\end{figure}

Figure~\ref{fig:statementsemantics} presents the semantics of statements.
We follow a big-step operational approach, where every statement updates the belief state as well as an optional true environment.
While the program execution only updates the belief state, we model programs as simultaneously nondeterministically updating the true environment to provide an easy specification of the set of legal observation inputs.
Namely, a legal observation is one that could have come from the true environment.
A null true environment signifies that, due to nondeterministic control flow, the true environment took a different branch than the belief state.
Observations under a null true environment are illegal.
The nondeterminism of the true environment impacts the belief state through the observations, but for fixed observation inputs, the belief state update is deterministic.
Every statement produces, given an initial belief state and true environment, a {\em configuration}, which is either a new belief state and new optional true environment or an error $\error$.
We augment the result with an observation list, which documents all observations and is an element of the grammar
\begin{align*}
    \ObsLst \; \bnf \; & \var : \num :: \ObsLst \; \mid \; \ncode{nil}
\end{align*}
In other words, an observation list is a list of associations of variable names to values.
We use the notation $\tuple{\bstate, \optpstate, \statement} \Downarrow \tuple{\config \; | \; \obslst}$ to mean that the belief state $\bstate$ and optional true environment $\optpstate$ produce the configuration $\config$ augmented with the observation list $\obslst$.

\paragraph{Assignment and Choose.} The meaning of either an assignment or choose statement is a new environment, if an original environment was present, and belief state with the statement variable rebound to a new value.
In the case of assignment, the value is given by evaluating the expression.
In the case of a choose statement the value must be consistent with the proposition, meaning that if we replace the placeholder $\ncode{.}$ with the new value, the proposition must hold.
In both cases, the new belief state is obtained by applying this process to each environment in the initial belief state.
A choose statement is nondeterministic with respect to the true environment, but deterministic with respect to the belief state.

\paragraph{Assertions.} The meaning of an assertion, if its predicate is true, is to return the input belief state and environment. 
If its predicate is false, the assertion returns an error.

\paragraph{Observations.} An observation $\lcode{observe} x$ does not modify the true environment.
It does modify the belief state to be consistent with the true environment on the observed variable $\var$ by only keeping those environments in the initial belief state that have the same value for $\var$ as in the true environment. 
The semantics also specify that the value of $\var$ is in the observation list, and that an error occurs if the true environment is null.

\paragraph{If Statements.} 
If the if statement's condition is deterministic (i.e., it is either true in all environments in the belief state or false in all environments), then the execution takes the appropriate branch.
This is specified by semantic rules that require $\bstate \vDash \Box \prop$ or $\bstate \vDash \Box \ncode{(!}\prop\ncode{)}$, where $\bstate$ is the initial belief state and $\prop$ is the if statement's condition.

If the statement's condition is nondeterministic (as specified by requiring $\bstate \vDash \Diamond p \code{\&\&} \Diamond \ncode{(!}p\ncode{)}$), the if statement executes both branches, sending as belief-state input to each the set of environments in which the condition has the appropriate value. 
It sends the true environment as input to the branch it actually takes and the null environment to the other branch.
The resulting belief state is then the union of environments resulting from either branch, and the resulting true environment is from the branch that the initial true environment actually took.

If an if statement's condition is nondeterministic, then neither of its branches can make observations, or else the result is an error.
This is because it is unclear what interaction with the true environment means within a branch that environment did not necessarily take.

\paragraph{Infer Statements.} The semantics of infer statements are similar to the semantics of if statements in an ordinary language, where infer operates solely on the belief state. 
If the belief state satisfies the condition, it evaluates the first branch and otherwise evaluates the second branch.

\paragraph{While Loops.} 
The semantics of while loops is defined recursively using if statements.
This is similar to a standard equivalence notion for while-loop programs (see \cite{winskel} Section 2.5).
We additionally include an assertion that requires the loop invariant to be true.

\paragraph{Composition and Skip.} The semantics of statement composition and skips are standard, except that they have been appropriately extended to include the observation list.
Sequencing concatenates observation lists using the $\concat$ operator.

\paragraph{Errors.} We have elided for clarity from Figure~\ref{fig:statementsemantics} the full semantics of how errors propagate.
We assume that errors propagate maximally throughout the program, so if at any point the semantics produce $\error$, the whole program produces $\error$.

\subsection{Semantic Properties}

In this section, we establish several properties of the semantics in Figure~\ref{fig:statementsemantics} that we posit any belief programming system should satisfy.
These properties constrain the semantics so that the belief state updates respect the true environment updates.

The first property is that beliefs should be sound. 
This means that for each environment in the belief state, and each new environment and new belief state that are reachable according to the semantics, the new environment is in the new belief state.
We formalize this as follows:

\begin{theorem}[Belief Soundness]{\ \\}
\vspace{-.35cm}
\begin{center}
    If $\pstate \in \bstate$ and $\tuple{\bstate, \pstate, \statement} \Downarrow \tuple{\bstate', \pstate' \; | \; o}$, then $\pstate' \in \bstate'$.
\end{center}
\vspace{-.3cm}
    \label{thm:bsound}
\end{theorem}

\begin{proof}
    By structural induction on derivations of $\Downarrow$.
    Details are in Appendix~\ref{sec:bsoundproof}.
\end{proof}
This means that it is impossible for the true environment to lie outside of the belief state. 

In addition, beliefs should be precise. 
This means that for every environment in a new belief state reachable from an initial belief state and the semantics, there is an environment in the initial belief state from which the new environment is reachable.
We formalize this as follows:

\begin{theorem}[Belief Precision]{\ \\}
\vspace{-.35cm}
\begin{center}
	If $\tuple{\bstate, \optpstate, \statement} \Downarrow \tuple{\bstate', \optpstate' \; | \; o}$, then for every
    $\pstate_\bstate' \in \bstate'$, there is some $\pstate_\bstate \in \bstate$ such that  $\tuple{\bstate, \pstate_\bstate, \statement} \Downarrow \tuple{\bstate', \pstate_\bstate' \; | \; o}$
\end{center}
\vspace{-.3cm}
\label{thm:bprecise}
\end{theorem}

\begin{proof} 
    By structural induction on derivations of $\Downarrow$.
    Details are in Appendix~\ref{sec:bpreciseproof}.
\end{proof} 
This means that every environment in the belief state could be the true environment.

Finally, beliefs should be deterministic given observations. 
This means that the belief state depends on the true environment only through observations.
We formalize this as follows:

\begin{theorem}[Belief Determinism]{\ \\}
\vspace{-.35cm}
\begin{center}
    If 
        $\tuple{\bstate, \optpstate_1, \statement} \Downarrow \tuple{\bstate_1', \optpstate_1' \; | \; o_1}$ and
        $\tuple{\bstate, \optpstate_2, \statement} \Downarrow \tuple{\bstate_2', \optpstate_2' \; | \; o_2}$ and
        $o_1 = o_2$, then $\bstate_1' = \bstate_2'$.
    \label{thm:bdeterm}
\end{center}
\vspace{-.3cm}
\end{theorem}

\begin{proof}
    By structural induction on derivations of $\Downarrow$ using a strengthened induction hypothesis.
    Details are in Appendix~\ref{sec:bdetermproof}.
\end{proof}

\subsection{Execution Model}

Here, we discuss how belief state updates execute in absence of the true environment.
This models how the belief program executes.
According to Theorem~\ref{thm:bdeterm}, the semantics in Figure~\ref{fig:statementsemantics} depends on the true environment $\optpstate$ only through the observation list $\obslst$.
By projecting out the operations on belief states, we can use the semantics in Figure~\ref{fig:statementsemantics} to compute the new belief state using only the initial belief state and the sequence of observations. 
In other words, if we define the belief execution $\Downarrow_\bstate$ as follows
\begin{mathpar}
    \inferrule
        {
            \tuple{\bstate,\optpstate, \statement} \Downarrow \tuple{\bstate', \optpstate' \; \mid \; \obslst}
        }
        {
            \tuple{\bstate, \obslst, \statement} \Downarrow_\bstate \bstate'
        }
\end{mathpar}
then $\Downarrow_\bstate$ is a partial function of $\bstate$, $\obslst$, and $\statement$.
The function $\Downarrow_\bstate$ gives the semantics of the belief program's execution on a concrete sequence of observations $\obslst$.

\section{Epistemic Hoare Logic}
\label{sec:logic}

\newcommand{\fresh}{\; \text{fresh in} \;}
\begin{figure}
    \begin{mathpar}
        \inferrule
            { 
                \var_0 \fresh \modalprop, \expr\\\\
                \modalprop' = \modalprop[\var_0/\var]\\
                \expr' = \expr[\var_0/\var]
            }
            {
                \htriplepc{\ndet}{\exists \hat{\qvar} . \; \modalprop}{\var \code{=} \expr}{
                    \exists \hat{\qvar}. \; \modalprop' \code{\&\&}
                    \Box \ncode{(} \var \code{==} \expr' \ncode{)}
                }
            }
            
        \inferrule
            {
                \htriplepc{\ndet}{\existsprop}{\statement}{\existsprop'}
            }
            {
                \htriplepc{\determ}{\existsprop}{\statement}{\existsprop'}
            }

        \inferrule
            { 
                \var_0 \fresh \modalprop, \prop\\\\
                \modalprop' = \modalprop[\var_0/\var]\\
                \prop' = \prop[\var_0/\var][\var/\ncode{.}]
            }
            {
                \htriplepc{\ndet}{\exists \hat{\qvar}. \; \modalprop}
                {\var \rcode{= choose(} \prop \ncode{)}}{
                    \exists \hat{\qvar}. \; \modalprop' \code{\&\&} 
                    \Box\ncode{(}\prop'\ncode{)}
                }
            }

        \inferrule
            { \forall \beta. \; \beta \vDash \existsprop \Rightarrow %
                                \beta \vDash \existsprop^a }
            {
                \htriplepc{\ndet}{\existsprop}{
                    \lcode{assert} \existsprop^a
                }{\existsprop}
            }

        \inferrule
            {
                \htriplepc{\pc}{\existsprop}{\statement_1}{\existsprop'}\\
                \htriplepc{\pc}{\existsprop'}{\statement_2}{\existsprop''}
            }
            {
                \htriplepc{\pc}{\existsprop}{\statement_1 \code{;} \statement_2}{\existsprop''}
            }

        \inferrule { }
            {
                \htriplepc{\ndet}{\existsprop}{\ncode{skip}}{\existsprop}
            }

        \inferrule
            { 
                \qvar_{n+1} \fresh \modalprop\\
                \modalprop' = \modalprop[\qvar_{n+1}/\var]
            }
            {
                \htriplepc{D}{\exists \qvar_0, \qvar_1, \dots, \qvar_n. \; \modalprop}
                {\lcode{observe} \var}{
                    \exists \qvar_0, \qvar_1, \dots, \qvar_n, \qvar_{n+1}. \; 
                        \modalprop' \code{\&\&}
                    \Box \ncode{(} \var \code{==} \qvar_{n+1} \ncode{)}
                }
            }

        \inferrule
            { 
                \htriplepc{\ndet}{\exists \hat{\qvar}. \; \ncode{(}\Box \prop \ncode{)}
                    \code{\&\&} \modalprop}{\statement_1}
                    {\exists \hat{y}'. \; \Box \prop'}\\
                \htriplepc{\ndet}{\exists \hat{\qvar}. \; \ncode{(}\Box \ncode{!}\prop \ncode{)}
                    \code{\&\&} \modalprop}{\statement_2}
                    {\exists \hat{y}'. \; \Box \prop'}
            }
            {
                \htriplepc{\ndet}{\exists \hat{\qvar}. \; \modalprop}
                {
                    \lcode{if} \prop \code{\{} \statement_1
                    \code{\} else \{} \statement_2 \rcode{\}}
                } {
                    \exists \hat{y}'. \; \Box \prop'
                }
            }

        \inferrule
            { 
                \forall \bstate. \; \bstate \vDash \exists \hat{\qvar}. \; \modalprop
                    \Rightarrow \bstate \vDash \Box \prop \code{||} 
                        \Box\ncode{(!} \prop \ncode{)}\\
                \htriplepc{\determ}{\exists \hat{\qvar}. \; \ncode{(}\Box \prop \ncode{)}
                    \code{\&\&} \modalprop}{\statement_1}
                    {\existsprop'}\\\\
                \htriplepc{\determ}{\exists \hat{\qvar}. \; \ncode{(}\Box \ncode{!}\prop \ncode{)}
                    \code{\&\&} \modalprop}{\statement_2}
                    {\existsprop'}
            }
            {
                \htriplepc{\determ}{\exists \hat{\qvar}. \; \modalprop}
                {
                    \lcode{if} \prop \code{\{} \statement_1
                    \code{\} else \{} \statement_2 \rcode{\}}
                } {
                    \existsprop'
                }
            }

        \inferrule
            { 
                \htriplepc{\pc}{\exists \hat{\qvar}. \; \modalprop^i 
                    \code{\&\&} \modalprop}{\statement_1}
                    {\existsprop'}\\
                \htriplepc{\pc}{\exists \hat{\qvar}. \; \ncode{!(} \modalprop^i \lcode{) \&\&}
                     \modalprop}{\statement_2}
                    {\existsprop'}
            }
            {
                \htriplepc{\pc}{\exists \hat{\qvar}. \; \modalprop}
                {
                    \lcode{infer} \modalprop^i \code{\{} \statement_1
                    \code{\} else \{} \statement_2 \rcode{\}}
                } {
                    \existsprop'
                }
            }
            
    \inferrule { 
            \htriplepc{\ndet}{\existsprop'}{\statement}{\existsprop''}\\
            \forall \bstate. \; \bstate \vDash \existsprop \Rightarrow
                \bstate \vDash \exists \hat{y}. \; \Box \prop^I\\
            \forall \bstate. \; \bstate \vDash \exists \hat{y}. \; 
                \Box \prop^I \code{\&\&} \Box p
                \Rightarrow \bstate \vDash \existsprop' \\
            \forall \bstate. \; \bstate \vDash \existsprop'' \Rightarrow
                \bstate \vDash \exists \hat{y}. \; \Box \prop^I 
        }
        {
            \htriplepc{\ndet}{\existsprop}
            {\lcode{while} \prop \code{\{} 
                \exists \hat{y}. \; \Box \prop^I \code{\} \{} \statement \code{\}}}
            { \exists \hat{y}.\; \Box \ncode{(!} \prop \code{\&\&} \prop^I \lcode{)} }
        }

        \inferrule { 
            \htriplepc{\determ}{\existsprop'}{\statement}{\existsprop''}\\\\
            \forall \bstate. \; \bstate \vDash \exists \hat{y}. \; \modalprop^I \Rightarrow
                \bstate \vDash \Box \prop \code{||} \Box \ncode{(!} \prop \ncode{)}\\
            \forall \bstate. \; \bstate \vDash \existsprop \Rightarrow
                \bstate \vDash \exists \hat{y}. \; \modalprop^I \\\\
            \forall \bstate. \; \bstate \vDash \exists \hat{y}. \; 
                \modalprop^I  \code{\&\&} \Box p
                \Rightarrow \bstate \vDash \existsprop' \\
            \forall \bstate. \; \bstate \vDash \existsprop'' \Rightarrow
                \bstate \vDash \exists \hat{y}. \; \modalprop^I 
        }
        {
            \htriplepc{\determ}{\existsprop}
            {\lcode{while} \prop \code{\{} 
                \exists \hat{y}. \; \modalprop^I \code{\} \{} \statement \code{\}}}
            { \exists \hat{y}.\; \Box \ncode{!(} \prop \lcode{) \&\&} \modalprop^I }
        }

        \inferrule {
            \htriplepc{\pc}{\existsprop'}{\statement}{\existsprop''}\\\\
            \forall \bstate. \; \bstate \vDash \existsprop \Rightarrow 
            \bstate \vDash \existsprop'\\\\
            \forall \bstate. \; \bstate \vDash \existsprop'' \Rightarrow \existsprop'''
        }
        {
            \htriplepc{\pc}{\existsprop}{\statement}{\existsprop'''}
        }
    \end{mathpar}
    \caption{Epistemic Hoare Logic rules. We use the notation $\htriplepc{\pc}{\existsprop}{\statement}{\existsprop'}$ to mean that in a context $\pc$ the statement $s$ maps belief states satisfying $\existsprop$ to new belief states satisfying $\existsprop'$.}
    \label{fig:logic}
\end{figure}

In this section, we present Epistemic Hoare Logic and sketch a proof of its soundness.
Figure~\ref{fig:logic} gives the inference rules for Epistemic Hoare Logic.
Each rule yields a deduction $\htriplepc{\pc}{\existsprop}{\statement}{\existsprop'}$, where the context $\pc$ is drawn from the grammar
$
    \pc \; \bnf \; \ndet \; \mid \; \determ
$.
The purpose of the context is to ensure observations do not occur under nondeterministic control flow, as that would result in an error according to the semantics.
The deduction $\htriplepc{\determ}{\existsprop}{\statement}{\existsprop'}$ meas that, assuming the statement $s$ is executed under deterministic control flow and terminates, $s$ maps belief states satisfying the pre-condition $\existsprop$ to new belief states satisfying the post-condition $\existsprop'$.
The deduction $\htriplepc{\ndet}{\existsprop}{\statement}{\existsprop'}$ means the same thing, but where the enclosing control flow may be nondeterministic.

\paragraph{Assignment and Choose.} Assignment and choose statements conjunct the pre-condition with a new proposition.
In the case of assignment, the new proposition encodes that the value of the variable is equal to the result of the expression.
In the case of a choose statement, the new proposition encodes that the value of the variable must be consistent with the choose statement's proposition.
In either case the previous value of the variable is encoded with a fresh variable.

\paragraph{Assertions.} An assertion has identical pre- and post-condition propositions. 
In order to apply the rule, the developer must show that the pre-condition implies the asserted proposition.

\paragraph{Observations.} An observation adds a new existentially quantified variable $\qvar_{n+1}$ to the variable list from the pre-condition. 
The post-condition ensures that the value of the observed variable is deterministic by stating that in every environment it is equal to $\qvar_{n+1}$.
Moreover, $\qvar_{n+1}$ is not completely unrestricted; it must satisfy all properties that the observed variable satisfied in the precondition.
Observations always require that the enclosing control flow is deterministic.

\paragraph{Composition and Skip.} The rules for statement sequencing and skip statements are standard; they are the same as in classical Hoare logic~\citep{floyd, hoare}

\paragraph{If and Infer.} Both if and infer statements require that both branches satisfy the same post-condition.
For if, the pre-condition of each branch includes the statement's condition under the $\Box$ modality.
For infer, the pre-condition of each branch includes the statement's condition, which is itself a modal proposition.

Furthermore, the post-condition of an if statement must use the $\Box$ modality, whereas the post-condition of an infer statement may be any existential proposition.
A more conventional approach would be to have both branches of the if statement imply the same predicate as the post-condition.
For this to be sound, we would need to show that if $\beta_1 \models p_\exists$ and $\beta_2 \models p_\exists$ then $\beta_1 \cup \beta_2 \models p_\exists$, which is in fact false.
However, it is true that if $\beta_1 \models \Box p$ and $\beta_2 \models \Box p$, then $\beta_1 \cup \beta_2 \models \Box p$.

To preserve a deterministic context, developers must ensure that the the if statement's condition has the same value in all environments.

\paragraph{While.} 
To verify a while loop with Epistemic Hoare Logic, the developer must prove several properties of the initial pre-condition $p_\exists$, the loop invariant $p_\exists^I$, the loop body's pre-condition $p_\exists'$, and the loop body's post-condition $p_\exists''$.

The developer must show that the pre-condition implies the loop invariant, the invariant implies the body's pre-condition, and the body's post-condition implies the invariant.
Implications such as these are a standard feature of proving properties using Hoare rules, although the overall proof rules are sometimes structured differently.

To preserve a deterministic context, developers must also ensure that the the while loop's condition has the same value in all environments.
\paragraph{Rule of Consequence.} The rule of consequence that any proposition that implies the pre-condition may be substituted for the pre-condition. Conversely, any proposition implied by the post-condition may be substituted for the post-condition.

\subsection{Soundness}

In this section, we formalize and establish the soundness of Epistemic Hoare Logic with respect to the semantics of \langname{}.
We start by explaining what supporting lemmas are needed, and then state the main theorem and sketch the proof.

\subsubsection{Substitution}

The main theorem depends on a number of lemmas that relate the substitutions in Figure~\ref{fig:logic} to the environment mapping in Figure~\ref{fig:statementsemantics}.

The following lemma gives the required property for expressions. 
It states that if we evaluate an expression under a new environment with a fresh variable $\var_0$ that is a rebinding of $\var$, we are free to rename $\var$ to $\var_0$ in the expression without changing the result.

\begin{lemma}[Expression Substitution]
{\ \\}
\vspace{-.35cm}
\begin{center}
    If $\var_0$ is fresh in $\expr$, then $\tuple{\pstate, \expr} \Downarrow \num \iff \tuple{\pstate[\var_0 \mapsto \pstate(\var)], \expr} \Downarrow \num$ and 
    $\tuple{\pstate, \expr} \Downarrow \num \iff \tuple{\pstate[\var_0 \mapsto \pstate(\var)], \expr[\var_0/\var]} \Downarrow \num$
\end{center}
\vspace{-.3cm}
\end{lemma}

\begin{proof}By structural induction on expressions.\end{proof}

By similar means, we establish analogous properties for propositions, modal propositions, and existential propositions.
The full set of lemmas is in Appendices \ref{sec:substproofs}-\ref{sec:constsubstproofs}.

\subsubsection{Observation List Emptiness}

The main theorem depends upon a lemma that states that observations cannot happen under nondeterministic control flow.
We have formalized this property as follows:

\begin{lemma}[Observation List Emptiness]{\ \\}
    \vspace{-.35cm}
    \begin{center}
    If $\htriplepc{\ndet}{\existsprop}{\statement}{\existsprop'}$ and $\tuple{\bstate, \optpstate, \statement} \Downarrow \tuple{\config \; | \; \obslst}$, then $\obslst = \ncode{nil}$.
    \end{center}
    \vspace{-.3cm}
\end{lemma}
\begin{proof}
    By structural induction on derivations of $\Downarrow$.
    The case of sequencing uses the fact that $\ncode{nil} \concat \code{nil} = \ncode{nil}$
\end{proof}

\subsubsection{Soundness Theorem}

In this section, we establish the soundness of the logic. Our approach is to show partial correctness, meaning that if the program terminates, its final state is described by the post-condition.

The theorem has two parts.
First, we establish the soundness of the logic in the case where control flow may be nondeterministic. The theorem states that if the belief state satisfies the pre-condition, and the semantics produces a configuration, the configuration includes a new belief state that satisfies the post-condition.
Second, we establish the soundness of the logic under deterministic control flow.
The theorem states that if the initial belief state satisfies the pre-condition, and the true environment is in the belief state, then the program executes without error and the new belief state satisfies the post-condition.
\begin{theorem}[Logic Soundness]{\ \\}
\vspace{-.35cm}
\begin{center}
    \begin{enumerate}
        \item If $\htriplepc{\ndet}{\existsprop}{\statement}{\existsprop'}$, $\bstate \vDash \existsprop$, and $\tuple{\bstate, \optpstate, s} \Downarrow \tuple{\config \; | \; \obslst}$, then $\config = (\bstate', \optpstate')$ and $\bstate' \vDash \existsprop'$.
        \item If $\htriplepc{\determ}{\existsprop}{\statement}{\existsprop'}$, $\sigma \in \bstate$, $\bstate \vDash \existsprop$, and $\tuple{\bstate, \pstate, s} \Downarrow \tuple{\config \; | \; \obslst}$, then $\config = (\bstate', \pstate')$ and $\bstate' \vDash \existsprop'$.
    \end{enumerate}
\end{center}
\vspace{-.2cm}
\label{thm:ehlsound}
\end{theorem}

\begin{proof} 
    \begin{enumerate}
        \item By structural induction on derivations of $\Downarrow$. The cases for assignments and choose statements rely on substitution lemmas. 
The cases for deterministic if, infer, and while follow from induction hypotheses, although while loops require destructing the semantic rule. 
For non-deterministic if statements, we show that 
$\bstate_1 \vDash \exists \hat{y}. \; \Box \prop \wedge 
\bstate_2 \vDash \exists \hat{y}. \; \Box \prop \Rightarrow
\bstate_1 \cup \bstate_2 \vDash \exists \hat{y}. \; \Box \prop 
$, which follows from standard principles of first-order logic.
        The details are in Appendix~\ref{sec:ehlsoundproof1}.
    \item By structural induction on derivations of $\Downarrow$. The cases except for$\code{observe}$are similar to those above, except that we use Theorem~\ref{thm:bsound} to ensure the premises of the induction hypotheses.
        The case for$\code{observe}$relies on the assumption that $\sigma \in \beta$ to instantiate $\qvar_{n+1}$. The details are in Appendix~\ref{sec:ehlsoundproof2}.
    \end{enumerate}
\end{proof}

\section{Case Study: The Mars Polar Lander}
\label{sec:casestudy}

\begin{figure}

    \begin{lstlisting}[basicstyle=\ttfamily\small]
//Controller Initialization
prev_td_1 = 0; cur_td_1 = 0; prev_td_2 = 0; cur_td_2 = 0;
health_1 = 1; health_2 = 1; 
engine_enabled = 1; event_enabled = 0;

while (engine_enabled == 1) {
    //Controller loop start
    input radar_alt;$\label{code:readradar}$
    
    prev_td_1 = cur_td_1; input cur_td_1;$\label{code:readtd1}$
    prev_td_2 = cur_td_2; input cur_td_2;$\label{code:readtd2}$
        
    state_1 = 0; //Missing, probable cause of crash.$\label{code:statestart}$
    state_2 = 0; //Missing, probable cause of crash.$\label{code:mplannii}$

    if (prev_td_1 == 1 && cur_td_1 == 1) {
        state_1 = 1
    };
    if (prev_td_2 == 1 && cur_td_2 == 1) {
        state_2 = 1
    };$\label{code:stateend}$
    if ((state_1 == 1 && health_1 == 1) || $\label{code:enginestart}$
        (state_2 == 1 && health_2 == 1) && 
        event_enabled == 1) {
        engine_enabled = 0;
    }$\label{code:engineend}$
    //Indicator health check
    if (radar_alt < 40 && event_enabled == 0) {$\label{code:healthstart}$
        if (prev_td_1 == 1 && cur_td_1 == 1) {
            health_1 = 0;
        };
        if (prev_td_2 == 1 && cur_td_2 == 1) {
            health_2 = 0;
        }
        event_enabled = 1;
    }$\label{code:healthend}$
}
    \end{lstlisting}
    \caption{Code of the Mars Polar Lander}
    \label{fig:mplcode}
\end{figure}

In this section, we show how belief programming and Epistemic Hoare Logic could be used to implement and verify the control software of the Mars Polar Lander (MPL). 
The MPL is a lost space probe, hypothesized to have crashed into the surface of Mars during descent due to a control software error~\citep{mpl}. 
We do not claim that belief programming is the first or only technique for preventing the loss of the MPL. 
However, the notoriety and subsequent investigation of the MPL's loss has resulted in ample documentation~\citep{mpl} useful for illustrating in detail how belief programming and Epistemic Hoare Logic work.

The code presented in this section is written in \langname{}, except that for convenience we define two pieces of syntactic sugar.
The syntax $\prop_1 \code{=>} \prop_2$ desugars to $\ncode{!}\prop_1 \code{||} \prop_2$
and the syntax $\var \code{=} \prop$ desugars to 
$\lcode{if} \prop \code{\{} \var \rcode{= 1}
\code{\} else \{} \var \rcode{= 0} \rcode{\}}$.

The code in Figure~\ref{fig:mplcode} is the piece of MPL's software~\citep{mpl} responsible for the final phase of its Martian descent. 
The code uses a radar altimeter as well as two touch sensors on the landing legs to monitor its progress along the descent. 
Note that this is a simplification from the original software, which used three touch sensors.
The code consists of a state estimator to determine when it reaches the Martian surface, and control code to shut off its engine once it does.

\paragraph{Reading Observations} On Line~\ref{code:readradar} the controller reads the value of the radar altimeter into the variable \lstinline{radar_alt}.
On Lines \ref{code:readtd1} and \ref{code:readtd2} the code reads the values of the touchdown sensors into \lstinline{cur_td_1} and \lstinline{cur_td_2}.
It also stores their previous values in \lstinline{prev_td_1} and \lstinline{prev_td_2}.

\paragraph{State Updates} The block of code from Line~\ref{code:statestart}-\ref{code:stateend} sets the state variables \lstinline{state_1} and \lstinline{state_2} based on the values of the touchdown sensors. 
Specifically, if an individual sensor has indicated a \lstinline{1} for two iterations in a row, its state is set to \lstinline{1}. 
Otherwise, its state is set to \lstinline{0}.

Notably the annotated lines (Lines \ref{code:statestart} and \ref{code:mplannii}) were missing from the original software, meaning that any two positive sensor readings in a row were sufficient to permanently set the state to \lstinline{1}.
It is hypothesized that this was part of the sequence of events that caused the MPL to crash, and these two lines are the recommended fix~\citep{mpl}.

\paragraph{Engine Shutdown} 
The block of code from Line~\ref{code:enginestart}-\ref{code:engineend} determines when to shut down the engine.
If the health check has completed (see below) and at least one of the healthy indicators registers a touchdown, then the program sets \lstinline{engine_enabled} to \lstinline{0}, shutting down the engine.

\paragraph{Health Check}
The block of code from Line~\ref{code:healthstart}-\ref{code:healthend} performs a health check on the touchdown indicators.
The health check assigns into the health variables \lstinline{health_0} and \lstinline{health_1}, and is performed the first time the radar altimeter indicates an altitude lower than 40 meters.
At this point, the lander is off of the ground so both touchdown indicators should read \lstinline{0}.
If an indicator reads \lstinline{1} on both the current and previous time step, it is assumed to be defective, and its health variable is set to \lstinline{0}.
After the health check completes, the program sets the flag \lstinline{event_enabled} to \lstinline{1} to allow the touchdown indicators to shut down the engine.

\subsection{Error Model}

From the above code, we can deduce several sources of error the MPL control software was designed to be robust with respect to:

\begin{itemize}
    \item {\it Transient False Positives.} If a touch sensor is momentarily triggered, the software will not immediately assume the lander has contacted the ground. 
        The software requires two time steps in a row where the sensor was positive before it sets its state variable to true and thereby registers that it has contacted the ground. 
        The assumption here is that the duration of the transient false positive is no longer than one time step.

    \item {\it Permanent False Positives.} If a touch sensor is defective, it may constantly send out an indication that the lander has contacted the ground. 
        This is detected and corrected for by the indicator health check on Lines~\ref{code:healthstart}-\ref{code:healthend}. 
        This block of code checks if the touch sensor yields a non-transient positive contact signal when the lander is just under 40 meters above the ground. 
        If so, the code assumes that sensor is defective and ignores its output for the engine shutoff decision. 
        The assumption here is that at most one sensor will be defective.

    \item {\it Permanent False Negatives.} If a touch sensor is defective, it may constantly send out an indication that the lander has not contacted the ground when in fact it has. 
        The above code accounts for this by using two sensors and shutting off the engine when either one indicates a touchdown. 
        The assumption here is that at most one sensor will be defective.
\end{itemize}

\paragraph{Landing Leg Deployment.} Another source of error for the MPL that is not obvious from the code above but that has been well-documented is landing leg deployment.
When the landing legs deploy about 1500 meters above the surface, the process can result in false positives that exceed the one time step assumed for transients~\cite{mpl}. 
Without the two annotated lines (Lines \ref{code:statestart} and \ref{code:mplannii}), this would cause the sensor's state variable to be permanently set to \lstinline{1}, causing the engine to shut down immediately after the health check completed.

\begin{figure}
    \begin{lstlisting}[basicstyle=\ttfamily\small]
//Model Initialization
prev_err_1 = 0; prev_err_2 = 0; trans_td = 0;
alt = 8000;$\label{code:altinit}$
time_on_ground = 0;
//Permanent errors
perm_1 = choose(. == 0 || . == 1);$\label{code:permstart}$
perm_2 = choose((. == 0 || . == 1) && (perm_1 == 1 => . == 0));
perm_1_v = choose(. == 0 || . == 1); 
perm_2_v = choose(. == 0 || . == 1);$\label{code:permend}$
//Controller initialization
...$\label{code:dots1}$
while(engine_enabled == 1) $\label{code:loopcondstart}$
{ (alt > 0 => engine_enabled == 1) && $\label{code:loopinvmpl}$
  (trans_td == 0 => time_on_ground < 2) }$\label{code:loopcondend}$
{
    //Model start
    if (alt == 0 && engine_enabled == 1) {$\label{code:togstart}$
        time_on_ground = time_on_ground + 1
    };$\label{code:togend}$
    alt_rate = choose(0 <= . && . <= 39 && . <= alt);$\label{code:altupdatestart}$
    alt = alt - alt_rate;
    radar_alt = choose (38 <= . - alt && . - alt <= 38);$\label{code:altupdateend}$

    err_1 = choose((prev_err_1 == 1 => . == 0) && (. == 0 || . == 1));$\label{code:errstart}$
    prev_err_1 = err_1;
    err_2 = choose((prev_err_2 == 1 => . == 0) && (. == 0 || . == 1));
    prev_err_2 = err_2;
    if (alt == 0 && (err_1 == 1 || err_2 == 1)) { trans_td = 1; };$\label{code:errend}$

    leg_err = 1400 <= alt && alt <= 1600;$\label{code:legdep}$
    if perm_1 { cur_td_1 = perm_1_v } $\label{code:td1start}$
    else if (leg_err == 1 || err_1 == 1) { 
        cur_td_1 = choose(. == 0 || . == 1)
    } else { cur_td_1 = alt == 0; };$\label{code:td1end}$
    
    if perm_2 { cur_td_2 = perm_2_v }$\label{code:td2start}$
    else if (leg_err == 1 || err_2 == 1)  {
        cur_td_2 = choose(. == 0 || . == 1)
    } else { cur_td_2 = alt == 0; };$\label{code:td2end}$

    //Model end
    //Controller loop start
    ...$\label{code:dots2}$
}
    \end{lstlisting}
    \caption{Model for verification of the Mars Polar Lander.}
    \label{fig:mplmodel}
\end{figure}

\subsubsection{Formalization} The nondeterministic program in Figure~\ref{fig:mplmodel} formalizes these sources of error. 
It provides inputs to the control software by setting the variables \lstinline{cur_td_1}, \lstinline{cur_td_2}, and \lstinline{radar_alt}.
It can be composed with the control software by inlining the code in Figure~\ref{fig:mplcode} on Lines \ref{code:dots1} and \ref{code:dots2}.
After composing with the control software, the resulting overall program could then be verified to prove the loop invariant on Lines \ref{code:loopinvmpl} and \ref{code:loopcondend} of Figure~\ref{fig:mplmodel} always holds.
We now describe each piece of the formal model in more detail.

\paragraph{Permanent Errors}

The block of code on Lines~\ref{code:permstart}-\ref{code:permend} of Figure~\ref{fig:mplmodel} models permanent errors, both false positive and false negative.
Each of the variables \lstinline{perm_1} and \lstinline{perm_2} is \lstinline{1} if its sensor has suffered a permanent failure, with the assumption being that neither of these two variables will be \lstinline{1} at the same time.
The variables \lstinline{perm_1_v} and \lstinline{perm_2_v} store the permanent error values, which are the constant values that each of the sensors will read when after suffering a permanent error.

\paragraph{Loop Condition}

The code on Lines~\ref{code:loopcondstart}-\ref{code:loopcondend} of Figure~\ref{fig:mplmodel} specifies the same loop condition as in Figure~\ref{fig:mplcode}, but also adds a loop invariant.
The invariant specifies that when the lander is off the ground, its engine is enabled.
Verifying this property would ensure that the software does not have the bug that caused the MPL to crash.
The invariant also specifies on Line~\ref{code:loopcondend} that the lander should spend less than 2 time steps on the ground with the engine on.
This is another constraint the MPL was designed to satisfy~\citep{mpl}.
However, the model in Figure~\ref{fig:mplmodel} admits transient false negative errors, which can violate this constraint in extreme cases.
Thus, on Line~\ref{code:loopcondend} of Figure~\ref{fig:mplmodel}, we assume this holds in the case where there are no transient false negatives.

\paragraph{Time on Ground}

The code on Lines~\ref{code:togstart}-\ref{code:togend} of Figure~\ref{fig:mplmodel} measures the amount of time the lander has spent on the ground with the engine on and stores it in \lstinline{time_on_ground}.
This is measured purely to evaluate the loop invariant and is not passed to the controller.

\paragraph{Altitude}

The code on Lines~\ref{code:altupdatestart}-\ref{code:altupdateend} of~\ref{fig:mplmodel} specifies how altitude changes and the error model of the radar altimeter.
It stores the rate of altitude change in \lstinline{alt_rate}, the new altitude in \lstinline{alt}, and the altimeter reading in \lstinline{radar_alt}.
We assume that the altitude changes by at most 39 meters and that the altimeter is accurate to within 38 meters.
The original motivation for including touchdown sensors on the MPL was that the radar altimeter is inaccurate below about 40 meters~\citep{mpl}.
This model is designed to conservatively capture this property while still ensuring that the condition on Line~\ref{code:healthstart} of Figure~\ref{fig:mplcode} triggers the indicator health check.

Furthermore, note that on line ~\ref{code:altinit} of Figure~\ref{fig:mplmodel} we have specified that the entry point to the program is when the lander is at an altitude of 8 kilometers.

\paragraph{Transient Errors} 
The code on Lines~\ref{code:errstart}-\ref{code:errend} of Figure~\ref{fig:mplmodel} models transient errors.
The variables \lstinline{err_1} and \lstinline{err_2} are set to be \lstinline{1} if a transient error occurred for the first or second touchdown sensor, respectively. 
The previous values of these variables are stored in \lstinline{prev_err_1} and \lstinline{prev_err_2}.
This code specifies that a transient error can occur for a sensor if it did not occur at the previous time step.
Furthermore, if a transient occurs after touchdown (i.e.\ a false negative), the code specifies that the variable \lstinline{trans_td} is set to \lstinline{1}.

\paragraph{Landing Leg Deployment}
To account for landing leg deployment errors, the model sets the flag \lstinline{leg_err} whenever the landing gear are deploying.
This occurs at about 1500 meters for the MPL~\cite{mpl}.  
We have modeled a deployment window of 100 meters around this nominal value, so that landing leg deployment occurs between 1400 and 1600 meters.

\paragraph{Touchdown Indicators}
The code on Lines~\ref{code:td1start}-\ref{code:td1end} of Figure~\ref{fig:mplmodel} specifies how the first touchdown indication is generated from the various sources of error.
The result is stored in \lstinline{cur_td_1}.
If the indicator suffered a permanent error, then it returns its permanent error value.
Otherwise, during landing leg deployment or a transient error, it may output either \lstinline{0} or \lstinline{1}.
If none of these errors are present, then it indicates \lstinline{1} iff the lander has touched the surface (i.e.\ the altitude is 0 meters).

The code on Lines~\ref{code:td2start}-\ref{code:td2end} of Figure~\ref{fig:mplmodel} specifies how the second touchdown indication is generated and is entirely symmetric.

\begin{figure}
    \begin{lstlisting}[basicstyle=\ttfamily\small]
// Model Initialization
...$\label{code:bpinline1}$
engine_enabled = 1;$\label{code:einit}$
while (engine_enabled == 1)$\label{code:mplbploopcond}$
{ $\Box$((alt > 0 => engine_enabled == 1) && $\label{code:mplbploopinv1}$
    (trans_td == 0 => time_on_ground < 2)) }$\label{code:mplbploopinv2}$
{
    // Model start
    ...$\label{code:bpinline2}$
    // Model end
    observe radar_alt;
    observe cur_td_1;
    observe cur_td_2;

    infer $\Box$(alt == 0) {$\label{code:mplinfer}$
        engine_enabled = 0
    }
}
    \end{lstlisting}
    \caption{Implementation of the Mars Polar Lander with belief programming}
    \label{fig:mplbp}
\end{figure}

\subsection{Belief Programming}
We now show how to use the model in Figure~\ref{fig:mplmodel} to construct a belief program to control the MPL.
We show a fragment of the belief program in Figure~\ref{fig:mplbp}.
This fragment can be completed by inlining the appropriate parts of Figure~\ref{fig:mplmodel} into Lines \ref{code:bpinline1} and \ref{code:bpinline2} of Figure~\ref{fig:mplbp}.

The belief program executes by observing each of the sensor readings generated from the model.
It then determines, on Line~\ref{code:mplinfer}, whether these are sufficient to guarantee the lander is on the ground.
If so, it shuts down the engine.
The belief program also modifies the loop invariant to have a $\Box$ modality, stating that it must be true in every environment (Lines \ref{code:mplbploopinv1} and \ref{code:mplbploopinv2}).

\subsection{Verification}

In this section, we will explain how to verify the loop invariant on Lines~\ref{code:mplbploopinv1}-\ref{code:mplbploopinv2} of Figure~\ref{fig:mplbp}.
We simplify the problem here by only considering the first condition on Line~\ref{code:mplbploopinv1}.
The remaining condition is considered in Appendix~\ref{sec:mplverification}.

\paragraph{Initialization Post-condition}

As specified by the rules in Figure~\ref{fig:logic}, we must show that the initialization code generates a post-condition that satisfies the loop invariant.
This post-condition can be written as $\Box\ncode{(engine\_enabled == 1)} \code{\&\&} \dots$,
where the $\Box$-proposition is generated by the code on Line~\ref{code:einit} of Figure~\ref{fig:mplbp}.
Now, we can apply the fact that
\[
    \forall \pstate. \; \pstate \vDash \ncode{(engine\_enabled == 1)} \Rightarrow 
    \pstate \vDash \ncode{(alt > 0 => engine\_enabled == 1)}
\]
and Theorems \ref{thm:lifting} and \ref{thm:k} to see that 
\[
\bstate \vDash \Box\ncode{(engine\_enabled == 1)} \code{\&\&} \dots \Rightarrow
\bstate \vDash \Box\ncode{(alt > 0 => engine\_enabled == 1)}
\]

\paragraph{Loop Body Post-condition}
According to the rules in Figure~\ref{fig:logic}, we must show that the loop body's post-condition implies the loop invariant.
We can summarize the post-condition as
\begin{align*}
    & \Box\ncode{(engine\_enabled\_0 == 1)} \code{\&\&} \\
    & \ncode{(}\Box\ncode{(alt == 0)} \code{\&\&} \Box\ncode{(engine\_enabled == 0)}\ncode{)} \code{||}\\
    & \ncode{(}\Diamond\ncode{(alt != 0)} \code{\&\&} \Box\ncode{(engine\_enabled == engine\_enabled\_0)}\ncode{)}
\end{align*}
where the first line comes from the loop condition on Line~\ref{code:mplbploopcond} and the second and third line come from the infer statement on Line~\ref{code:mplinfer}.
We assume we have applied the standard rule for strongest-postcondition predicates and taken the disjunction of each branch of the \lstinline{infer}~\citep{floyd}.

We can now show the post-condition implies the invariant. The disjunction gives us two cases. In the first, we can assume that $\Box\ncode{(alt == 0)}$.
Now, we can apply the fact that by vacuous truth,
\[
    \forall \pstate. \; \pstate \vDash \ncode{(alt == 0)} \Rightarrow 
    \pstate \vDash \ncode{(alt > 0 => engine\_enabled == 1)}
\]
and Theorems \ref{thm:lifting} and \ref{thm:k} to see that 
\[
\bstate \vDash \Box\ncode{(alt == 0)} \Rightarrow
\bstate \vDash \Box\ncode{(alt > 0 => engine\_enabled == 1)}
\]
In the second case, we follow a similar logic to the initialization post-condition argument above, with the additional premise that $\Box\ncode{(engine\_enabled == engine\_enabled\_0)}$.

\section{Implementation}
\label{sec:impl}

We have demonstrated the feasibility of a belief programming implementation that directly represents the belief state with a set.
Specifically, we wrote the implementation in C using a hash set data structure to represent beliefs.

Called CBLIMP, our implementation is a shallow embedding of BLIMP into C; i.e.\ it is a C library that implements each of the core BLIMP primitives as a C function.
CBLIMP's \texttt{observe} function takes, in addition to the current belief state and the variable to be observed, a parameter that is the observed value of the variable.
CBLIMP's \texttt{infer} function returns a boolean to be used as a branching condition by the C program, and
CBLIMP's \texttt{if} function takes as parameters callbacks for each branch that execute on modified belief states.
All regular C variables in a CBLIMP program can be considered to be deterministic with respect to the belief state (i.e.\ they have the same value in all environments in the belief state).

CBLIMP includes functions that extend BLIMP primitives to simulate the true environment via random sampling.
The simulated \texttt{observe} function presents a different interface: instead of taking the observed value as a parameter, it uses the value from the simulated true environment.

We enforce that environments are finite by augmenting \texttt{choose} statements to include a range that the newly assigned variable must belong to.
For example, as part of the MPL model, Line~\ref{code:errstart} of Figure~\ref{fig:mplmodel} gives the choose statement
\begin{center}
\texttt{err\_1 = choose((prev\_err\_1 == 1 => . == 0) \&\& (. == 0 || . == 1));}
\end{center}
This specifies that \texttt{err\_1} is a boolean value (i.e. it is either 0 or 1) and that whenever \texttt{prev\_err\_1} is 1, \texttt{err\_1} must be 0.
In our implementation, we have alternatively specified this as
\begin{center}
\texttt{err\_1 = choose(0,1, prev\_err\_1 == 1 => . == 0);}
\end{center}
where the arguments \texttt{0} and \texttt{1} of the \texttt{choose} statement are the lower and upper bounds of the range that \texttt{err\_1} must belong to. Note that these statements are equivalent; whereas in the original statement we specified that \texttt{err\_1} is a boolean using the choose statement's proposition, in the new statement we have instead specified this using the range bounds.

\paragraph{Research Question.} We evaluated CBLIMP to answer the question: does the direct implementation achieve practical performance given the latency requirements of the domain?
For the UAV example, a step latency of 1 second is required to match the latency of common GPS receivers~\cite{sparkfungps}. 
For the MPL example, a latency of 10ms is required~\cite{mpl}.

\paragraph{Benchmarks.} We used the UAV example with a 100-step time horizon and the MPL example as benchmarks. The code is nearly identical to that in the paper, with the following changes:
\begin{itemize}
    \item The MPL benchmark includes an additional intermediate variable that captures the error between the true altitude and the radar altitude.
    \item We manually determined bounds for every variable specified by a choose statement to facilitate use of our implementation's augmented choose statements.
\end{itemize}

We also implemented another version of MPL that uses a different \emph{grid size}.
Note that the original version of MPL defines a uniform discretization grid for both true and radar altitude at a resolution of 1 meter.
We modified this to be a non-uniform grid on both true and observed altitude with higher-altitude grid cells exponentially larger than lower-altitude grid cells.
We call this benchmark "MPL-Exp". We have provided its \langname{} source code and verified it in Appendix~\ref{sec:mplexp}.

\paragraph{Methodology.} We ran each benchmark 5 times and recorded the mean and standard deviation of the time to execute a single iteration of the main while loop. 
We also recorded the maximum time taken by an iteration across all runs.

To construct observations to send as inputs to the belief program, we simulated the true environment alongside the belief program's execution, sampling the new true environment uniformly at random when we encountered a choose statement.
While the latency measurements include both the time to update the belief state and run the simulation, we expect them to be dominated by belief state operations.

\begin{wraptable}{l}{0.5\textwidth}
    \caption{Results of performance experiments}
    \label{tab:results}
    \begin{tabular}{ | l | r | r | }
        \hline
        Benchmark & Mean +- Std. Dev. & Maximum \\ \hline
        UAV & 2.49 +- 0.37 ms & 4.41 ms \\
        MPL & 2.45 +- 1.42 s & 11.9 s \\
        MPL-Exp & 0.76 +- 0.56 ms & 2.37 ms \\\hline
    \end{tabular}
\end{wraptable}

\paragraph{Results.} Table~\ref{tab:results} summarizes the step latency for each benchmark.
We can see that with the UAV benchmark, the direct implementation of belief programming is practical in the sense that the step latency is well under the 1s threshold.
However, with the MPL example, the direct implementation is not practical as-is.
The latency requirement of 10ms is about 1000x faster than the worst-case latency of our implementation.
By contrast, our modified "MPL-Exp" benchmark is practical for the MPL problem because its worst-case latency is considerably below the 10ms threshold.

\paragraph{Threats to validity.}
We ran these benchmarks on a 2017 MacBook Pro with an i7-7920HQ CPU at 3.10GHz and 16 GB of 2133 MHz DDR3.
Both of our benchmarks operate in domains that necessitate embedded computers that are less powerful.
The UAV example enjoys a comfortable margin over the required latency; it is likely that the processors common on larger drones could meet the latency requirements.
The MPL-Exp benchmark would typically be run on a small embedded processor, both for reliability benefits and because such processors are typically the ones hardened against cosmic radiation.
Although we expect such a processor to be slower than a modern laptop computer, with a comfortable 5x slowdown margin until it violates the latency requirement, we speculate that this benchmark could meet the requirement with standard performance engineering techniques.

\section{Future Work}

\subsection{Implementation Efficiency}
There is an open question of how to design an efficient implementation for the belief programming runtime.
CBLIMP directly implements the semantics of Figure~\ref{fig:statementsemantics} using an exhaustive representation of belief states. 
This implementation scales poorly with the number of variables in the program, and we discuss here more efficient potential runtime implementation approaches.

\paragraph{Runtime SMT} One approach could symbolically execute \langname{} constructs to collect constraints that an SMT solver would use to evaluate modal propositions. 
This approach would make use of the enhanced performance of SMT solvers compared to an exhaustive approach, and bears similarities to other languages that deploy solvers at runtime (e.g.~\cite{jeeves, planb}).

\paragraph{Restricted belief states} Known efficient implementations exist when the belief states admitted by the language are more restricted than the full powerset of environments.
Examples of restricted classes of belief states studied in the literature include ellipsoids~\cite{ellipsoids} and polytopes~\cite{polytopes}.

\paragraph{Synthesis} Using the semantics of Figure~\ref{fig:statementsemantics} as a specification, a variety of synthesis tools~\cite{cegis, vsa, dedsynth} exist that could potentially generate more efficient implementations than naive enumeration.
The goal would be to translate \texttt{infer} statements to ordinary \texttt{if} statements, using synthesis to construct a predicate for the \texttt{if} statement that is a function of the observed values and is semantically equivalent to the \texttt{infer} statement's predicate.

\subsection{Logic Automation}

The logic in Figure~\ref{fig:logic} and the theorems in Section~\ref{sec:proptheorems} enable sound, manual reasoning about the behavior of belief programs.
As with many program logics, the gap from manual to automated reasoning is the need for automated techniques for invariant inference and implication checking.

\paragraph{Invariant Inference.} As in traditional program logics, a while loop requires a loop invariant. 
Although propositions in the Epistemic Hoare Logic include modalities, classic approaches such as template-based invariant inference may be directly applicable~\citep{houdini} via templates that include modalities.
An additional distinct difference from many traditional program logics is that the rules for if statements require developers to manually determine a suitable post-condition or, in other words, provide an invariant.
Here too template-based techniques may be directly applicable.
In either case, there may also be new opportunities for analysis-based invariant inference techniques that account for modalities.

\paragraph{Discharging Implications.} 
A classic approach to discharge implications that appear in the premises of Hoare logic rules is to employ an automatic theorem prover such as Z3~\citep{z3}.
To apply this approach to Epistemic Hoare Logic, one would need to contend with the modalities in propositions.
Recent work holds out the promise of automated reasoning techniques for modal implications via a reduction to SMT~\citep{modalsmt, modalsat}.

\section{Related Work}
\label{sec:related}

\paragraph{Set-based Uncertainty.} 
\cite{signalsets} is a survey paper that gives an overview of how set-based uncertainty is used in the signal processing domain. It explains how programs can over-approximate the true belief state (using e.g. ellipsoids~\cite{schweppe}; more recent work has studied polytopes~\cite{polytopes}), and how the quality of approximation can be measured~\cite{schweppe} to determine if the resulting belief state is too large. It also gives efficient algorithms~\cite{kaczmarz, cimmino} for a restricted set of operations on approximate belief states . By contrast, belief programming reasons about the exact belief state and provides a richer set of operations. However, it cannot achieve the same computational efficiency as an approximation.

\paragraph{Classical Verification}
In Sections~\ref{sec:example} and \ref{sec:casestudy}, we alluded to how the UAV and MPL examples could be verified using classical techniques.
Here, we explain this process in more detail.

The developer would first compose their handwritten environment model (Figures~\ref{fig:dronemodel} and \ref{fig:mplmodel}) with their handwritten state estimator (Figures~\ref{fig:dronecode}  and \ref{fig:mplcode}).
The resulting program is in the language IMP~\cite{winskel} with the addition of choose statements that provide nondeterminism.
The developer could obtain a Hoare logic for this language by either extending the logic of IMP~\cite{winskel} to include choose statements, or by rewriting it to a language such as GCL~\cite{gcl} which supports nondeterminism natively and also has a Hoare logic.
Finally, the developer would apply the Hoare logic to the program, which requires discharging verification conditions. Because the proposition language is the standard propositional calculus, we expect there are many techniques in the literature that the developer could apply to this end.

The advantage of classical verification, relative to belief programming, is that developers can expect to draw on a wide variety of verification literature to solve the problem, as the formulation is relatively standard. 
The disadvantage is that developers must write the raw code of the state estimator by hand, which is a tedious process. 
By contrast, in belief programming, the state estimator consists of \texttt{infer} statements, which are more intuitive to use.

\paragraph{Epistemic and Belief Revision Logics.} The Epistemic Hoare Logic we present in Section~\ref{sec:logic} is similar to dynamic Epistemic logics such as public announcement logic~\citep{pal} and action-based logics~\citep{del}.
Whereas these logics typically use either propositions or abstract action spaces to modify the belief state, our logic uses a belief program to do so.

\paragraph{Synthesis.} To avoid writing state estimators by hand, developers might generate a state estimator by synthesizing it directly from the environment model~\citep{cegis, vsa, dedsynth}.
Such a problem would be challenging due to having hidden state encoded in the belief program that must be explicit in the state estimator.

For example, in the desired handwritten state estimator in Figure~\ref{fig:mplcode}, the program contains the variables \texttt{prev\_td\_1} and \texttt{prev\_td\_2}, which are not related to any of the variables in the model in Figure~\ref{fig:mplmodel}.
We anticipate that existing synthesis tools will have difficulty automatically inferring the existence of such hidden variables.

\paragraph{Dynamic Constraint Solving.} Some existing programming languages~\citep{jeeves,planb} perform constraint solving at runtime using an SMT solver.
Such approaches are necessarily similar to a belief program, which also represents a constrained set of program states at runtime.
Existing systems were designed for other domains, and do not articulate complete set of \texttt{choose}, \texttt{observe}, and \texttt{infer} constructs that \langname{} has.

\subsection{Relationship to Probabilistic Programming}

Probabilistic programming languages (PPLs) are also designed to enable developers to reason about uncertainty. We compare BLIMP to PPLs on three core axes: language features, contemporary reasoning techniques, and the practicality of inference (i.e., the implementation of \texttt{infer}) . 

\paragraph{Language Features} BLIMP's primary programming constructs are \texttt{choose}, \texttt{observe}, and \texttt{infer} and have probabilistic analogs in probabilistic programming languages. BLIMP's \texttt{choose} has the classic interpretation of non-probabilistic, nondeterministic choice. The analog in probabilistic programming is the probabilistic \texttt{sample} construct that randomly samples a value according to a distribution. BLIMP's \texttt{observe} has a similar semantics to \texttt{observe} constructs in PPLs.  BLIMP's \texttt{infer}, which enables the program itself to perform inference, has some support in PPLs as well~\citep{staton,probzelus}.

In general, probabilistic programming provides a more flexible modeling mechanism than BLIMP's set-based uncertainty, enabling a developer to specify distributions for nondeterministic outcomes.  For applications for which probabilistic models of outcomes are available, probabilistic programming can be an appropriate choice. However, probabilistic models of outcomes are not available for all applications (e.g., our MPL application) and introducing distributions can complicate reasoning, as we discuss below. BLIMP is an additional design point for applications that do not necessarily benefit from probabilistic modeling. 

\paragraph{Reasoning} The verification of probabilistic programs is an active area of research~\cite{passert, probverif}. This work typically provides the ability to express and verify the probability that an assertion is true of the program. In principle, it is possible to verify BLIMP-like modal assertions in such a framework. Specifically, $\Box$ maps to an assertion that a proposition is true with probability 1; $\Diamond$ maps to an assertion that a proposition is true with positive probability. 

However, a significant challenge with reasoning about probabilistic programs is that the distribution over states at any given point in a program may not have an analytical characterization. Specifically, the composition of a standard, well-known probability distribution (e.g., a Gaussian distribution) with a computation can result in a distribution that is not well-characterized by a standard, well-known distribution for which standard statistical quantities (such as mean and variance) are easily accessible. 

Therefore the full modeling, programming, and reasoning workflow must carefully consider the distributions used in \texttt{sample} statements such that they adhere to the application's uncertainty model and that the resulting distributions in the rest of the program can be precisely reasoned about at an acceptable level of complexity.

BLIMP is, instead, an additional design point for modeling uncertainty that need not rely on an appropriate selection of \texttt{sample} distributions or, more generally, the complexity of techniques for reasoning about probabilistic constructs. 

\paragraph{Practicality of Inference.} BLIMP's runtime implementation to support inference -- i.e.\ \texttt{infer} -- tracks all possible environments. The resulting implementation is sound. A probabilistic programming language can take a similar strategy  -- i.e., exact inference -- if it desires sound inference.

However, probabilistic programming languages can also leverage approximate inference algorithms for probabilistic programs such as Sequential Monte Carlo methods~\cite{doucet-smc} that need only track a high-probability subset of the possible environments. These methods are approximate in that the probability of an event is estimated with a fidelity that is a function of the size and diversity of the tracked state.  Designing algorithms to select this state efficiently is application-specific. Contemporary diagnostics for Monte Carlo methods (e.g.\ \cite{ess, mcmcdiag}) are typically not sound in that they can indicate a good approximation when the true approximation is poor. Establishing bounds on the quality of the resulting approximation is still an active area of research~\cite{issample}.

Therefore, while approximate inference methods can in practice provide empirically good results, reasoning about the soundness of their results is still an active area of research.

\section{Conclusion}
\label{sec:conclusion}

In this paper, we presented belief programming and Epistemic Hoare Logic.
Belief programming enables developers to write programs that can be directly executed to give state estimators that are derived from environment models. Epistemic Hoare Logic enables developers to reason about belief programs. 
We discussed both by reference to the \langname{} language, with belief programming described by \langname{}'s semantics and Epistemic Hoare Logic operating over \langname{} statements.
We determined that belief programming is feasible by evaluating our BLIMP implementation, CBLIMP.
Taken together, this work lays new foundations for soundly reasoning about the behavior of software that executes in partially-observable environments.

%% Acknowledgments
\begin{acks}                            %% acks environment is optional
                                        %% contents suppressed with 'anonymous'
  %% Commands \grantsponsor{<sponsorID>}{<name>}{<url>} and
  %% \grantnum[<url>]{<sponsorID>}{<number>} should be used to
  %% acknowledge financial support and will be used by metadata
  %% extraction tools.
  We would like to thank Alex Renda, Deokhwan Kim, Ben Sherman, Jesse Michel, Cambridge Yang, Jonathan Frankle, and anonymous reviewers for their helpful comments and suggestions.
    This work was supported in part by the Office of Naval Research (ONR-N00014-17-1-2699).
Any opinions, findings, and
  conclusions or recommendations expressed in this material are those
  of the author and do not necessarily reflect the views of the
  Office of Naval Research.

%  This material is based upon work supported by the
%  \grantsponsor{GS100000001}{National Science
%    Foundation}{http://dx.doi.org/10.13039/100000001} under Grant
%  No.~\grantnum{GS100000001}{nnnnnnn} and Grant
%  No.~\grantnum{GS100000001}{mmmmmmm}.  
\end{acks}

%% Bibliography
\bibliography{belief-paper_arxiv}

%% Appendix
\newpage
\appendix
\section{Proofs}
\label{sec:proofs}
\subsection{Theorem~\ref{thm:bsound}: Belief Soundness}
\label{sec:bsoundproof}

Proceed by structural induction on derivations of $\Downarrow$. 
Individual cases are as follows:
\begin{itemize}
    \item {\bf Nondeterministic If Statements.} We case split on which branch the true environment takes.
If it is the true branch, then $\pstate \in \bstate_T$.
Then, by induction hypothesis $\pstate' \in \bstate'_T \Rightarrow \pstate' in \bstate'_T \cup \bstate'_F$.
The case for the false branch is symmetric.

    \item {\bf All other Statements.} This follows directly from the definition of the new belief state and the induction hypotheses.
\end{itemize}

\subsection{Theorem~\ref{thm:bprecise}: Belief Precision}
\label{sec:bpreciseproof}

Proceed by structural induction on derivations of $\Downarrow$.
In each case, we identify the choice of $\pstate_\bstate$ given $\pstate_\bstate'$ that satisfies the property.

\begin{itemize}
    \item {\bf Assignment.} Choose $\pstate_\bstate \in \bstate$ such that $\tuple{\pstate_\bstate, e} \Downarrow \num_\bstate$ and $\pstate_\bstate' = \pstate_\bstate[\var \mapsto \num_\bstate]$.
    \item {\bf Choose.} Choose $\pstate_\bstate \in \bstate$ such that 
        there exists a $\num_\bstate$ such that 
        $\pstate_\bstate \vDash \prop[\num_\bstate/.]$ and
        $\pstate_\bstate' = \pstate_\bstate[\var \mapsto \num_\bstate]$.
    \item {\bf Observe.} Choose $\pstate_\bstate = \pstate_\bstate'$.
    \item {\bf Nondeterministic If.} Case split on whether $\pstate_\bstate'$ is in $\bstate_T'$ or $\bstate_F'$.
        If it is in $\bstate_T'$, choose $\optpstate_T' = \pstate_\bstate$ and apply the inductive hypothesis for $s_1$.
        Otherwise, apply the inductive hypothesis for $s_2$.
        In either case, the resulting $\pstate_\bstate$ must be in $\bstate$.
    \item {\bf Other Statements.} Apply inductive hypotheses.
\end{itemize}

\subsection{Theorem~\ref{thm:bdeterm}: Belief Determinism}
\label{sec:bdetermproof}

First, we define the following \emph{prefix property}:
\begin{property}[Prefix]
    If $\tuple{\bstate, \pstate, \statement} \Downarrow \tuple{\bstate' \; | \; \obslst}$
    and $\tuple{\bstate, \pstate, \statement} \Downarrow \tuple{\bstate' \; | \; \obslst'}$
    where $\obslst \neq \obslst'$, then neither is $\obslst$ a prefix of $\obslst'$ nor $\obslst'$ a prefix of $\obslst$.

\end{property}

The prefix property states that if a belief program can generate two different observation lists, the lists have to fundamentally differ and cannot simply be extensions of each other.

Proceed by structural induction on derivations of $\Downarrow$.
We strengthen the inductive hypothesis with the prefix property.
Specific cases are as follows:
\begin{itemize}
    \item {\bf Sequencing.} We consider here the rule that produces non-\texttt{nil} observation lists through the concatenate operator.

        First, we show the prefix property. If $\obslst_1$ is different between the two executions, then the resulting concatenated list must satisfy the prefix property according to the mechanics of the concatenate operator.
        Otherwise, we apply the determinism inductive hypothesis on $s_1$ to show that the intermediate belief state $\bstate'$ is the same between executions.
        This means we can directly apply the prefix property inductive hypothesis on $s_2$ to show the result.
       
        Now, we show belief determinism.
        We can assume the length of all observation lists in this rule are the same across both executions.
        Otherwise, $\obslst_1$ in one execution would be a prefix of $\obslst_1$ in the other execution, violating the prefix property.
        Due to the mechanics of the list concatenate operation, this means we can assume the values of the lists are the same across both executions.
        Thus, we can apply the determinism inductive hypotheses to show the result.

    \item {\bf Observe.} The list always has length 1, which ensures the prefix property. 
        The new belief state is a function of the original belief state and the observed value, which ensures belief determinism.

    \item {\bf Other cases.} In the remaining cases, either the observation list is constant and the new belief state is a function of the original one, or the conclusion follows directly from induction hypotheses.
\end{itemize}

\subsection{Logic}

\subsubsection{Substitution Lemmas}
\label{sec:substproofs}

Here, we restate and reprove the substitution lemma from Section~\ref{sec:logic} and state and prove the other substitution lemmas we will need.

\begin{lemma}[Expression Substitution]
    If $\var_0$ is fresh in $\expr$, then $\tuple{\pstate, \expr} \Downarrow \num \iff \tuple{\pstate[\var_0 \mapsto \pstate(\var)], \expr} \Downarrow \num$ and 
    $\tuple{\pstate, \expr} \Downarrow \num \iff \tuple{\pstate[\var_0 \mapsto \pstate(\var)], \expr[\var_0/\var]} \Downarrow \num$
\end{lemma}

\begin{proof}
    By structural induction on the language of expressions.
\end{proof}

\begin{lemma}[Proposition Substitution]
    If $\var_0$ is fresh in $\prop$, then $\pstate \vDash \prop \iff \pstate[\var_0 \mapsto \pstate(\var)] \vDash \prop$ and 
    $\pstate \vDash \prop \iff \pstate[\var_0 \mapsto \pstate(\var)] \vDash \prop[\var_0/\var]$
\end{lemma}

\begin{proof}
    By structural induction on the language of propositions. The base case employs the expression substitution lemma above.
\end{proof}

In the following definition, we use the notation $\bstate[\var_0 \mapsto \var]$ to mean a belief state $\bstate'$ such that 
$\bstate' = \{ \pstate[\var_0 \mapsto \pstate(x)] \; | \; \pstate \in \bstate\}$

\begin{lemma}[Modal Proposition Substitution]
    If $\var_0$ is fresh in $\modalprop$, then $\bstate \vDash \modalprop \iff \bstate[\var_0 \mapsto \var] \vDash \modalprop$ and 
    $\bstate \vDash \modalprop \iff \bstate[\var_0 \mapsto \var] \vDash \modalprop[\var_0/\var]$
\end{lemma}

\begin{proof}
    By structural induction on the language of propositions. 
    The base cases employ the proposition substitution lemma above.
\end{proof}

\subsubsection{Agreement}
\label{sec;agreeproofs}

In this section, we show some \emph{agreement lemmas} that are also required to show the main soundness result.

\begin{lemma}[Expression Agreement]
    If $\sigma_1$ agrees with $\sigma_2$ on every location except $\var_0$ and, $\expr$ does not depend on $\var_0$, then $\tuple{\sigma_1, \expr} \Downarrow \num \iff \tuple{\sigma_2, \expr} \Downarrow \num$.
\end{lemma}

\begin{proof}
    By structural induction on the language of expressions.
\end{proof}

\begin{lemma}[Proposition Agreement]
    If $\sigma_1$ agrees with $\sigma_2$ on every location except $\var_0$ and, $\prop$ does not depend on $\var_0$, then $\sigma_1 \vDash \prop \iff \sigma_2 \vDash \prop$.
\end{lemma}

\begin{proof}
    By structural induction on the language of propositions.
    The base case follows from the expression agreement lemma above.
\end{proof}

\begin{lemma}[Modal Proposition Agreement]
    If $\bstate_1$ agrees with $\bstate_2$ on every location except $\var_0$ 
    (i.e.\ $\sigma_1 \in \bstate_1$ iff $\sigma_2 \in \bstate_2$ and $\sigma_2$ agrees with $\sigma_1$ on every location except $\var_0$), then
    $\bstate_1 \vDash \modalprop \iff \bstate_2 \vDash \modalprop$

\end{lemma}

\begin{proof}
    By structural induction on the language of modal propositions.
    The base case follows from the proposition agreement lemma above, and we also use the fact that belief agreement is symmetric to prove the only if case.
\end{proof}

\subsubsection{Constant Substitution}
\label{sec:constsubstproofs}

The following is an additional lemma used in the proof of the soundness of Epistemic Hoare logic:

\begin{lemma}[Expression Constant Substitution]
    If $\sigma(\var) = \num_\sigma$, then $\tuple{\sigma, \expr} \Downarrow \num$ iff $\tuple{\sigma, \expr[\num_\sigma/\var]} \Downarrow \num$.
\end{lemma}

\begin{proof}
    By structural induction on the language of expressions.
\end{proof}

\begin{lemma}[Proposition Constant Substitution]
    If $\sigma(\var) = \num_\sigma$, then $\sigma \vDash \prop$ iff $\sigma \vDash \prop[\num_\sigma/\var]$.
\end{lemma}

\begin{proof}
    By structural induction on the language of propositions.
    For the base cases, we apply the expression constant substitution lemma above.
\end{proof}

\begin{lemma}[Modal Proposition Substitution]
    If for all $\sigma \in \beta$, $\sigma(\var) = \num$, and there exists a $\sigma \in \beta$ such that $\sigma(\var) = \num$, then $\beta \vDash \existsprop \iff \beta \vDash \existsprop[\var/\num]$.
\end{lemma}

\begin{proof}
    By structural induction on the language of modal propositions.
    For the base cases, we apply the proposition constant substitution lemma above
\end{proof}

\subsubsection{Proof of Theorem~\ref{thm:ehlsound}: Part 1}
\label{sec:ehlsoundproof1}

Proceed by structural induction on derivations of $\Downarrow$.
Specific cases are as follows:
\begin{itemize}
    \item {\bf Assignment.} The first part of the post-condition follows from the agreement and substitution lemmas for modal propositions. 
        The second part follows from expression substitution.
    \item {\bf Choose.} The first part of the post-condition follows from the agreement and substitution lemmas.
        Then, we can show that since $\prop[\var_0/\var]$ does not depend on $\var$,
        $\pstate \vDash \prop[\var_0/\var][./\num] \iff 
         \pstate[\var \mapsto \num] \vDash \prop[./\var]$. 
        This can be shown by structural induction, similarly to the proof for the agreement and substitution lemmas.
        The second part of the post-condition follows from this fact.
    \item {\bf Nondeterministic If.}  
        By definition, $\bstate_T \vDash
                \exists \hat{\qvar}. \; \ncode{(}\Box \prop \ncode{)}
                    \code{\&\&} \modalprop$ and 
                $\bstate_F \vDash 
                \exists \hat{\qvar}. \; \ncode{(}\Box \ncode{!}\prop \ncode{)}
                    \code{\&\&} \modalprop$
        After applying inductive hypotheses, we can use the fact that $\bstate_T \vDash \Box \prop \wedge \bstate_F \vDash \Box \prop \Rightarrow \bstate_T \cup \bstate_F \vDash \Box \prop$ to show the overall result.

    \item {\bf Other Cases.} The other cases follow from applying inductive hypotheses and inlining definitions. 
        In the case of while loops, the premises must be destructed, and the extra premises of the logic rule ensure that the semantics do not evaluate to $\bot$.
\end{itemize}

\subsubsection{Proof of Theorem~\ref{thm:ehlsound}: Part 2}
\label{sec:ehlsoundproof2}

Proceed by structural induction on derivations of $\Downarrow$.
Most cases follow the same reasoning as part 1 above, and we cover here the cases that are unique to part 2.

\begin{itemize}
    \item {\bf Observe.} Choose $\qvar_{n+1} = \pstate(\var)$. 
        We show the first part of the post-condition by applying the constant substitution lemma for modal propositions. 
        The first premise of this lemma follows from the definition of $\bstate'$, and the second condition follows from the assumption that $\pstate \in \bstate$.
        The second part of the post-condition follows directly from the definition of $\bstate'$.
    \item {\bf Subtyping.} Because this part of the theorem is a strengthened version of the first part, any derivation produced by subtyping must be sound.
    \item {\bf Other Cases.} The result follows from applying inductive hypotheses.
\end{itemize}

\section{Theorems About Propositions}
\label{sec:proptheorems}

In addition to the logical rules in Figure~\ref{fig:logic}, developers need to prove the implications in the premises of those rules.
We propose that developers do so by lifting propositional reasoning to the modal operators.
The theorems in this section give a general set of reasoning tools for performing this lifting.

First, we show that modal operator $\Box$ commutes with conjunctions:

\begin{theorem}[$\Box$ commutes with $\rcode{\&\&}$]
    $\beta \vDash \Box \ncode{(} \prop_1 \code{\&\&} \prop_2 \ncode{)} \iff
    \beta \vDash \Box \prop_1 \code{\&\&} \Box \prop_2$
    \label{thm:box-and}
\end{theorem}

\begin{proof}From first-order logic instantiation.\end{proof}

We can similarly show that modal operator $\Box$ partially commutes with disjunctions, but the implication only applies in one direction

\begin{theorem}[$\Box$ commutes with $\rcode{||}$]
    $\beta \vDash \Box \prop_1 \code{||} \Box \prop_2 \Rightarrow
    \beta \vDash \Box \ncode{(} \prop_1 \code{||} \prop_2 \ncode{)}$
    \label{thm:box-or}
\end{theorem}

\begin{proof}By case analysis on the premise. Follows from the fact that $\sigma \vDash \prop_1 \Rightarrow \sigma \vDash \prop_1 \code{||} \prop_2$.\end{proof}

The $\Box$ and $\Diamond$ operators are also related by negation, in the same way that $\forall$ and $\exists$ are related in first-order logic. 
This means that, while we present all theorems in this section as applying to $\Box$, there are analogous theorems that apply to $\Diamond$.

\begin{theorem}[$\Box$-$\Diamond$ duality]
    $\beta \vDash \Box p \iff \beta \vDash \ncode{!} \Diamond \ncode{!} p$
    \label{thm:box-diamond}
\end{theorem}

\begin{proof}Follows from definitions and $\forall$-$\exists$ duality.\end{proof}

Next, we show two theorems that enable developers to push implications through modal operators. 
The first states that the $\Box$ modality applies to any formulas that are true across all environments.

\begin{theorem}[Knowledge of Theorems]
    $\Big( \forall \sigma. \; \sigma \vDash \prop_1 \Rightarrow \sigma \vDash \prop_2 \Big)
    \Rightarrow \Big( \beta \vDash \Box \ncode{(!} \prop_1 \code{||} \prop_2 \ncode{)} \Big)$
    \label{thm:lifting}
\end{theorem}

\begin{proof}Follows from $\beta$ being a subset of all possible $\sigma$.\end{proof}

The second theorem states that implications under $\Box$ can be lifted to modal propositions.

\begin{theorem}[Knowledge Instantiation]
    $\beta \vDash \Box \ncode{(!} \prop_1 \code{||} \prop_2 \ncode{)} 
    \Rightarrow \Big( \beta \vDash \Box \prop_1 \Rightarrow \beta \vDash \Box \prop_2 \Big)$
    \label{thm:k}
\end{theorem}

\begin{proof} Follows from first-order instantiation and transitivity of $\Rightarrow$.\end{proof}

Finally, we show that for propositions that do not depend on environment variables (i.e.\ they only depend on quantified variables and constants), truth in one environment implies truth in any.

\begin{theorem}[Environment Independence]{\ \\}
\vspace{-.4cm}
\begin{center}
    If $\prop$ does not contain any $\var$, then 
    $\beta \vDash \exists \hat{y}. \; \modalprop \code{\&\&} \Diamond \prop \Rightarrow 
    \beta \vDash \exists \hat{y}. \; \modalprop \code{\&\&} \Box \prop$
\end{center}
\vspace{-.3cm}
    \label{thm:envindep}
\end{theorem}

\begin{proof}By structural induction, we can show that $\sigma \vDash p$ is independent of $\sigma$. Thus, if $p$ is true, it is true under all environments.\end{proof}

\section{Verification of the UAV Example}
\label{sec:exampleverification}
This section is structured as follows.
Section~\ref{sec:bpverifpost} explains how propositions are generated from belief programs, with an emphasis on how the rules of Epistemic Hoare logic differ from the rules of classical Hoare logic.
Section~\ref{sec:bpverifinv} gives examples of proving an implication with modal propositions.
In each section, we illustrate concepts in the context of verifying the invariant preservation property of the loop body in Figure~\ref{fig:dronebp}.

\subsubsection{Post-condition}
\label{sec:bpverifpost}

\begin{figure}
\begin{lstlisting}
$\exists$ y.
  $\Box$(450 <= alt_0 && alt_0 <= 550) && $\label{code:postloopinv}$
  $\Box$(alt_0 - 25 <= alt_1 && alt_1 <= x_0 + 25) &&$\label{code:altchoose}$
  $\Box$(alt_1 - 25 <= y && y <= alt_1 + 25) &&$\label{code:obschoose}$
  $\Box$(obs == y) &&$\label{code:obsprop}$
  ( ($\Diamond$(alt_1 < 450) && $\Box$(cmd == 50)) ||$\label{code:disstart}$
    (!$\Diamond$(alt_1 < 450) && $\Diamond$(alt_1 > 550) && $\Box$(cmd == -50)) ||
    (!$\Diamond$(alt_1 < 450) && !$\Diamond$(alt_1 > 550) && $\Box$(cmd == 0))$\label{code:disend}$
  ) &&
  $\Box$(alt == alt_1 + cmd)$\label{code:altprop}$
\end{lstlisting}
\caption{Post-condition of the belief program at the end of the loop.}
\label{fig:dronepost}
\end{figure}

Epistemic Hoare logic is most closely related to strongest-postcondition logics such as~\citet{floyd} that generate a post-condition given a pre-condition and a program.
This means there is a deduction
\[
    \htriplepc{\determ}{\Box\ncode{(450 <= alt \&\& alt <= 550)}}{s}{\existsprop'}
\]
where the post-condition $\existsprop'$ is generated fairly directly by the program $s$.
Figure~\ref{fig:dronepost} shows such an $\existsprop'$, and here we explain how it corresponds to $s$, the loop body from Figure~\ref{fig:dronebp}.

\paragraph{Loop Invariant} The proposition on Line~\ref{code:postloopinv} of Figure~\ref{fig:dronepost} corresponds to the loop invariant on Line~\ref{code:bploopinv} of Figure~\ref{fig:dronebp} that we assumed as our pre-condition.
Because the variable $\ncode{alt}$ is reassigned later in the loop body, the logic renames it to $\ncode{alt\_0}$, a fresh variable that represents the previous value of $\ncode{alt}$ at the start of the loop.
Note that this differs from the standard approach to name conflicts, which uses existential quantification~\citep{floyd}.
This is because Epistemic Hoare logic needs to preserve the original quantification with $\Box$ and/or $\Diamond$ when referring to variables in the environment.

\paragraph{Choose Statements} The choose statement on Line~\ref{code:bpchoosei} of Figure~\ref{fig:dronebp} generates the proposition on Line~\ref{code:altchoose} of Figure~\ref{fig:dronepost}.
This proposition states that the previous altitude $\ncode{alt\_0}$ is within a distance of $25$ of the new altitude $\ncode{alt\_1}$, and is generated from the choose statement's proposition by replacing the placeholder $\code{.}$ with the new altitude.
The updated altitude is renamed to the fresh variable $\ncode{alt\_1}$ because of the reassignment of $\ncode{alt}$ on Line~\ref{code:altasgn} of Figure~\ref{fig:dronebp}.

The choose statement for $\ncode{obs}$ on Line~\ref{code:bpchooseii} of Figure~\ref{fig:dronebp} similarly generates the proposition on Line~\ref{code:obschoose} of Figure~\ref{fig:dronepost}, though as we explain next, the new value of $\ncode{obs}$ is renamed to the existentially quantified variable $\ncode{y}$.

\paragraph{Observations} The observation on Line~\ref{code:bpobserve} of Figure~\ref{fig:dronebp} generates the existentially quantified variable $\ncode{y}$.
This variable stands for the input observed value of $\ncode{obs}$, and thus must satisfy any constraints that $\ncode{obs}$ was originally under.
In this case, that means that $\ncode{y}$ must be within a distance of $25$ of the altitude $\ncode{alt\_1}$.
Furthermore, because $\ncode{obs}$ is being observed, we know that every environment must have it bound to its true value $\ncode{y}$.
Thus, the observation generates the proposition on Line~\ref{code:obsprop} of Figure~\ref{fig:dronepost}.

\paragraph{Infer} The infer statement on Lines~\ref{code:inferstart}-\ref{code:inferend} of Figure~\ref{fig:dronebp} generates the disjunction on Lines~\ref{code:disstart}-\ref{code:disend} of Figure~\ref{fig:dronepost}.
Each term in the disjunction corresponds to one of the branches of the infer statement.
Each term itself is a conjunction of the predicate that causes the branch to be taken and a proposition describing the actions of the branch.
In every case, the predicate is a combination of (possibly negated) $\Diamond$-propositions and the branch specifies that the value of $\ncode{cmd}$ is some constant in every environment.

\paragraph{Assignment}

The assignment on Line~\ref{code:altasgn} of Figure~\ref{fig:dronebp} generates the proposition on Line~\ref{code:altprop} of Figure~\ref{fig:dronepost}. This proposition specifies that in any environment in the belief state, the new altitude $\ncode{alt}$ is equal to the previous altitude $\ncode{alt\_1}$ plus the command $\ncode{cmd}$.

\subsubsection{Modal Implications}
\label{sec:bpverifinv}

In this section, we show how implications of ordinary propositions can be lifted to implications of modal propositions. 
We further demonstrate how this can be used to show that the invariant $\existsprop'$ presented in Figure~\ref{fig:dronepost} and derived in the Section~\ref{sec:bpverifpost} implies the loop invariant in Figure~\ref{fig:dronebp}.
This completes the proof that the program in Figure~\ref{fig:dronebp} preserves its loop invariant.

The proof that $\existsprop'$ implies the loop invariant is structured as follows.
The first step is to treat term of the disjunction on Lines~\ref{code:disstart}-\ref{code:disend} of Figure~\ref{fig:dronepost} as a separate case.
For clarity, we only discuss the first case on Line~\ref{code:disstart} in this section.
In this case, in addition to the post-condition, we can assume the belief state satisfies the proposition $\Diamond\ncode{(alt\_1 < 450)} \code{\&\&} \Box\ncode{(cmd == 50)}$. Our approach will be to a) show that this implies a $\Diamond$-proposition about only $\ncode{y}$ b) that we can convert this $\Diamond$-proposition to a $\Box$-proposition, and c) that the resulting $\Box$-proposition implies the loop invariant.

\paragraph{Implied $\Diamond$-Proposition} The first technique for lifting proposition implications to modal implications is that if for any environment $P \Rightarrow Q$ where $P$ and $Q$ are propositions over environments, then in any belief state $\Diamond P \Rightarrow \Diamond Q$. 
In this case, we take $P$ to be the proposition $\ncode{alt\_1 < 450}$ and $Q$ to be the proposition $\ncode{alt\_1 < y - 25 || y <= 475}$. 
This means we can assume that \lstinline[breaklines=true]{$\Diamond$(alt_1 < y - 25 || y <= 475)} is true of our belief state.
Furthermore, by the proposition on Line~\ref{code:obschoose}, we know that every environment in the belief state satisfies the negation of $\ncode{alt\_1 < y - 25}$, meaning that if an environment satisfies $Q$ it must be because it satisfies $\ncode{y <= 475}$.
Thus, we can assume that $\Diamond\ncode{(y <= 475)}$.

\paragraph{Conversion to $\Box$-proposition} Note that $\Box$ and $\Diamond$ quantify over environments, and the proposition $\ncode{y <= 475}$ depends only on the quantified variable $\ncode{y}$ and not on any variables in the environment. 
This means that if the proposition is true in some environment, it must be true in every environment.
Thus, we can assume $\Box\ncode{(y <= 475)}$

\paragraph{Implied $\Box$-Proposition} Another technique for lifting proposition implications to modal implications is that if for any environment $P \Rightarrow Q$, then for any belief state $\Box P \Rightarrow \Box Q$.
We now apply this to the propositions in Figure~\ref{fig:dronepost} and the assumption $\Box\ncode{(y <= 475)}$.
In this case, we take $P$ to be
\[
    \ncode{425 <= alt\_1 \&\& alt\_1 <= 575 \&\& alt\_1 - 25 <= y \&\& y <= alt\_1 + 25 \&\& y <= 475}
\]
and $Q$ to be $\ncode{475 <= alt\_1 + 50 \&\& alt\_1 + 50 <= 550}$. By similar logic, we can see that the proposition $\Box\ncode{(y <= 475)}$ combined with the propositions on Lines ~\ref{code:postloopinv}-\ref{code:obschoose} implies $\Box P$ and that $\Box Q$ combined with the proposition on Line~\ref{code:altasgn} implies the loop invariant.

\section{Verifying the MPL with Epistemic Hoare logic}
\label{sec:mplverification}
In this section, we will show a piece of the verification process of the MPL belief program in Figure~\ref{fig:mplbp}. 
Specifically, we will show how to verify the belief program's while loop preserves the loop invariant.
Note that some of the variable names in this section are slightly different than the names in Section~\ref{sec:casestudy}.

First, we strengthen the loop invariant to the following proposition.
This strengthened condition contains the additional properties:
\begin{itemize}
    \item {\it Altitude and time-on-ground nonnegativity.} The variables \lstinline{altitude} and \lstinline{time_on_ground} never go below \lstinline{0}.
    \item {\it Permanent Error Single-upset.} It is never the case that both touchdown sensors suffer permanent failures.
    \item {\it Existential Variables.} There are two existentially quantified variables that stand in for the observed touchdown sensor values. This means that if there are no transients on touchdown and the landing legs are not deploying, these sensors will properly indicate a touchdown.
    \item {\it Engine Disabled Soundness.} On the second time step after touchdown, if there are no transients on touchdown, the controller will disable the engine.
\end{itemize}
\begin{lstlisting}
$\exists$ y1, y2.
    $\Box$(altitude >= 0) &&
    $\Box$(time_on_ground >= 0) &&
    $\Box$(altitude > 0 => time_on_ground == 0) &&
    $\Box$(!(permanent_1 == 1 && permanent_2 == 1)) &&
    $\Box$(altitude > 0 => engine_enabled = 1) &&
    $\Box$((((y1 != 1 && permanent_1 != 1) ||
        (y2 != 1 && permanent_2 != 1)) && 
       transient_on_touchdown == 0 && 
       landing_leg_deployment == 0) => altitude != 0) ) &&
    $\Box$((transient_on_touchdown == 0 && time_on_ground == 1) => 
        engine_enabled == 0) &&
    $\Box$(transient_on_touchdown == 0 => time_on_ground < 2) &&
\end{lstlisting}
The strengthened invariant can be shown to be correct on the first iteration of the loop applying our logic to the model initialization code. 
Our goal here is to show that this invariant is preserved by the belief program's control loop.

Calling this proposition $\existsprop^I$, we can propagate it through the loop body using the rules in Figure~\ref{fig:logic}. 
For the post-conditions of each if statement, we assume a natural proposition that includes a disjunction representing both branches and wraps the disjunction in a $\Box$ modality.
Similarly, for each infer statement, we include a disjunction that uses the modality of the infer statement's condition.
This yields the following formula, which we call $\existsprop^\mathrm{MPL}$. We will not explain the contents of $\existsprop^\mathrm{MPL}$ at length, but will refer to it piecemeal in the remainder of the proof.

\begin{lstlisting}[xleftmargin=2em,numbers=left]
$\exists$ y_radar, y_td1, y_td2, y1, y2.
    $\Box$(altitude_0 >= 0) &&$\label{code:altznonneg}$
    $\Box$(time_on_ground_0 >= 0) &&$\label{code:tognonneg}$
    $\Box$(altitude_0 > 0 => time_on_ground_0 == 0) &&$\label{code:alttogz}$
    $\Box$(!(permanent_1 == 1 && permanent_2 == 1)) &&$\label{code:perm}$
    $\Box$(altitude_0 > 0 => engine_enabled_0 = 1) &&$\label{code:ezenabled}$
    $\Box$((((y1 != 1 && permanent_1 != 1) ||$\label{code:yassump}$ 
        (y2 != 1 && permanent_2 != 1)) && 
       transient_on_touchdown == 0 && 
       landing_leg_deployment_0 == 0) => altitude_0 != 0) ) &&
    $\Box$((transient_on_touchdown_0 == 0 && time_on_ground_0 == 1) => 
        engine_enabled_0 == 0) &&$\label{code:togezdis}$
    $\Box$(transient_on_touchdown_0 == 0 => time_on_ground_0 < 2) &&$\label{code:togbound}$
    ( $\Box$(altitude_0 == 0 && engine_enabled_ == 1 && $\label{code:togasgn}$
        time_on_ground == time_on_ground_0 + 1) ||
      (!(altitude_0 == 0 && engine_enabled_0 == 1) && 
        time_on_ground == time_on_ground_0)
    ) &&
    $\Box$(0 <= altitude_rate && altitude_rate <= 39 && $\label{code:altrate}$
        altitude_rate <= altitude_0) &&$\label{code:altitudenonneg}$
    $\Box$(altitude == altitude_0 - altitude_rate) &&$\label{code:altitudeasgn}$
    $\Box$((prev_error_1_0 == 1 => error_1 == 0) && 
        (error_1 == 0 || error_1 == 1)) &&
    $\Box$(prev_error_1 == error_1) &&
    $\Box$((prev_error_2_0 == 1 => error_2 == 0) && 
        (error_2 == 0 || error_2 == 1)) &&
    $\Box$(prev_error_2 == error_2) &&
    $\Box$((transient_on_touchdown == 0 && altitude == 0) =>$\label{code:totasgn}$
        (error_1 != 1 && error_2 != 1)) &&
    ( $\Box$(1400 <= altitude && altitude <= 1600 && 
        landing_leg_deployment == 1) ||
      ((1400 > altitude || 1600 < altitude) && 
        landing_leg_deployment == 0)
    ) &&
    $\Box$((permanent_1 == 1 && y_td1 == permanent_1_val) ||$\label{code:tdiasgn}$
      (permanent_1 != 1 && 
        (landing_leg_deployment == 1 || error_1 == 1) && 
            y_td1 == 0 || y_td1 == 1
      ) ||
      (permanent_1 != 1 && landing_leg_deployment != 1 && 
        error_1 != 1 && 
            altitude == 0 => y_td1 == 1 &&
            altitude != 0 => y_td1 == 0
      )
    ) &&
    $\Box$((permanent_2 == 1 && y_td2 == permanent_2_val) ||$\label{code:tdiiasgn}$
      (permanent_2 != 1 && 
        (landing_leg_deployment == 1 || error_2 == 1) && 
            y_td2 == 0 || y_td2 == 1
      ) ||
      (permanent_1 != 1 && landing_leg_deployment != 1 && 
        error_2 != 1 && 
            altitude == 0 => y_td1 == 1 &&
            altitude != 0 => y_td1 == 0
      )
    ) &&
    $\Box$(40 <= y_radar - altitude && y_radar - altitude <= 40) &&$\label{code:radaracc}$
    $\Box$(y_radar == radar_altitude) &&$\label{code:yradar}$
    $\Box$(y_td1 == current_touchdown_indicator_1) &&
    $\Box$(y_td2 == current_touchdown_indicator_2) &&
    ( ($\Box$(altitude == 0) && $\Box$(engine_enabled == 0)) ||$\label{code:edisabled}$
      ($\Diamond$(altitude != 0) && $\Box$(engine_enabled == engine_enabled_0))$\label{code:eenabled}$
    )
\end{lstlisting}
We will now show that 
$\forall \bstate. \; \bstate \vDash \existsprop^\mathrm{MPL} 
    \Rightarrow \bstate \vDash \existsprop^I$
We will proceed by handling each conjunction in $\existsprop^I$ individually.

\subsubsection{Altitude nonnegativity}

Here, we will show that 
$\forall \bstate.\; \bstate \vDash \existsprop^\mathrm{MPL} \Rightarrow
    \bstate \vDash \Box\ncode{(altitude >= 0)}$
We rely on the fact that, based on standard properties of numerical comparisons and arithmetic operators,
$\forall \pstate. \; \pstate \vDash \code{altitude\_rate <= altitude\_0} \Rightarrow
    \pstate \vDash \rcode{altitude\_0 - altitude\_rate >= 0}$
Thus, by applying Theorem~\ref{thm:lifting}, Theorem~\ref{thm:k}, and the assumptions on Lines \ref{code:altitudenonneg} and \ref{code:altitudeasgn}, we have the result.

\subsubsection{Time on ground non-negativity}

Here, we will show that
$\forall \bstate.\; \bstate \vDash \existsprop^\mathrm{MPL} \Rightarrow
    \bstate \vDash \Box\ncode{(time\_on\_ground >= 0)}$
From Lines \ref{code:tognonneg} and \ref{code:togasgn} of $\existsprop^\mathrm{MPL}$, we can deduce that
\begin{lstlisting}[columns=flexible]
$\Box$(time_on_ground_0 >= 0) &&
$\Box$(time_on_ground == time_on_ground_0 || time_on_ground == time_on_ground_0 + 1)
\end{lstlisting}
We will now analyze the two cases of the disjunction separately.
In the first case we use the following implication, which must be true by a simple substitution argument.
\begin{align*}
    \forall \pstate.\; \pstate & \vDash \code{(time\_on\_ground\_0 >= 0 \&\& %
        time\_on\_ground == time\_on\_ground\_0)} \Rightarrow \\
        \pstate & \vDash \code{time\_on\_ground >= 0}
\end{align*}
In the second case, we can use following implication, which follows from the interation of $\code{>=}$ and $\rcode{+}$.
\begin{align*}
    \forall \pstate.\; \pstate & \vDash \code{(time\_on\_ground\_0 >= 0 \&\& %
        time\_on\_ground == time\_on\_ground\_0 + 1)} \Rightarrow \\
        \pstate & \vDash \code{time\_on\_ground >= 0}
\end{align*}
Combining these theorems together and then applying Theorems \ref{thm:lifting} and \ref{thm:k} shows that \lstinline[breaklines=true]{$\Box$(time_on_ground >= 0)}.

\subsubsection{Single upset for permanent errors}

Here, we will show that
\lstinline[breaklines=true]{$\Box$(!(permanent_1 == 1 && permanent_2 == 1))}.
This is directly assumed on Line~\ref{code:perm} of $\existsprop^\mathrm{MPL}$.

\subsubsection{Engine enabled soundness}

Here, we will show that 
$\forall \bstate. \; \bstate \vDash \existsprop^\mathrm{MPL}
    \Rightarrow \bstate \vDash \Box \ncode{(altitude > 0 => engine\_enabled == 1)}$.
Notably, the original MPL software did not satisfy this condition and as a result likely shut off its engine too early resulting in a crash.

We case split on the disjunction on Line~\ref{code:edisabled}. 
In the first case on Line~\ref{code:edisabled}, the assumption that \lstinline{$\Box$(altitude == 0)} means that any environment in the belief state will vacuously satisfy \lstinline{altitude > 0 => engine_enabled == 1}.

In the second case on Line~\ref{code:edisabled}, we can assume that
\lstinline[breaklines=true]{$\Diamond$(altitude != 0) => $\Box$(engine_enabled == 1)}.
Inlining definitions and using Theorem~\ref{thm:box-diamond}, we can see that
\[
    \Box\ncode{(altitude == 0)} \code{||} \Box\ncode{(engine\_enabled == 1)}
\]
Then, applying Theorem~\ref{thm:box-or}, we can deduce that 
\[
    \Box\ncode{(altitude == 0 || engine\_enabled == 1)}
\]
which is equivalent to 
\[
    \Box\ncode{(altitude != 0 => engine\_enabled == 1)}
\]
Combining this with the fact that \lstinline[breaklines=true]{$\Box$(altitude >= 0)} Using Theorems \ref{thm:lifting} and \ref{thm:k} proves the result.

\subsubsection{Time on ground vs. altitude}
\label{sec:togalt}

Here, we will show that \lstinline[breaklines=true]{$\Box$(altitude > 0 => time_on_ground == 0)}.
The assumptions on Lines \ref{code:altznonneg}, \ref{code:alttogz}, \ref{code:ezenabled}, \ref{code:togasgn}, and \ref{code:altitudeasgn} mean that any environment in the belief state satisfies,
\begin{lstlisting}[columns=flexible]
altitude_0 > 0 => time_on_ground_0 == 0 &&
altitude_0 > 0 => engine_enabled == 1 &&
altitude_0 ==  0 => altitude == 0 &&
(altitude_0 == 0 ||
engine_enabled_0 != 1 || 
altitude_0 > 0 && time_on_ground == time_on_ground_0)
\end{lstlisting}
Thus, any environment in the belief state satisfies \lstinline[breaklines=true]{altitude_0 > 0 => time_on_ground == 0}

\subsubsection{Existential Variables}

In this section, we will show that
\begin{align*}
    \bstate \vDash \existsprop^\mathrm{MPL} \Rightarrow
    \bstate \vDash \exists & \code{y1, y2. ((y1 || y2) \&\&}\\
    & \code{transient\_on\_touchdown == 0 \&\& landing\_leg\_deployment == 0) =>}\\
    & \code{altitude == 0}
\end{align*}
The conditions on Lines \ref{code:totasgn}, \ref{code:tdiasgn}, and \ref{code:tdiiasgn} can be rearranged to show that
\begin{align*}
    \bstate \vDash \existsprop^\mathrm{MPL} \Rightarrow
    \bstate \vDash \exists  & \code{y\_td1, y\_td2.} \Box\lcode{(transient\_on\_touchdown == 0 \&\&}\\
    & \code{landing\_leg\_deployment == 0 \&\&}\\
    &\code{!(permanent\_1 == 1 \&\& permanent\_2 == 1) =>}\\
    & \code{((y\_td1 || y\_td2) <=> altitude == 0))}
\end{align*}
Because by the assumption on Line~\ref{code:perm} we have that the belief state satisfies \lstinline[breaklines=true]{$\Box$(!(permanent_1 == 1 && permanent_2 == 1))}, we can $\alpha$-rename \lstinline{y_td1} to \lstinline{y1} and \lstinline{y_td2} to \lstinline{y2} which implies the result.

\subsubsection{Engine Disabled Soundness}
\label{sec:engdissound}

In this section we will show that
\[
\forall \bstate. \; \bstate \vDash \existsprop^\mathrm{MPL} \Rightarrow
\bstate \vDash \Box \ncode{(time\_on\_ground == 2 => engine\_enabled == 0)}
\]
We case split on the disjunction on Line~\ref{code:eenabled}.
In the first case, we can assume that the belief state satisfies $\Box\ncode{(engine\_enabled == 0)}$ which ensures the result.

In the second case, we can assume that the belief state satisfies $\Diamond\ncode{altitude != 0}$. 
We note the following implication
\begin{align*}
    \forall \pstate . \; \pstate \vDash & \code{altitude != 0} \Rightarrow\\
    \pstate \vDash & \code{y\_radar > 1000 || } \\
                   & \code{((y1 != 1 || y\_td1 != 1) \&\& (y2 != 1 || y\_td2 != 1)) ||} \\
                   & \code{!( (((y1 == 1 \&\& y\_td1 == 1) || (y2 == 1 \&\& y\_td2 == 1)) \&\& y\_radar <= 1000)}\\
                   & \code{=> altitude == 0)}
\end{align*}
Next, we lift this implication to operate over $\Diamond$.
By taking the contrapositive of Theorem~\ref{thm:k} and applying Theorem~\ref{thm:box-diamond} we can show that 
$
(\forall \pstate. \; \pstate \vDash \prop_1 \Rightarrow \pstate \vDash \prop_2) 
\Rightarrow (\forall \bstate. \; \bstate \vDash \Diamond \prop_1 \Rightarrow \bstate \vDash \Diamond \prop_2)
$.
Applying this, we see that
\begin{align*}
    \forall \bstate . \; \bstate \vDash & \code{altitude != 0} \Rightarrow\\
    \bstate \vDash & \; \Diamond \ncode{(y\_radar > 1000) || } \\
                   & \; \Diamond \ncode{((y1 != 1 || y\_td1 != 1) \&\& (y2 != 1 || y\_td2 != 1)) ||} \\
                   & \; \Diamond \ncode{!( (((y1 == 1 \&\& y\_td1 == 1) || (y2 == 1 \&\& y\_td2 == 1)) \&\& y\_radar <= 1000)}\\
                   & \code{=> altitude == 0)}
\end{align*}
We start by addressing the final term of the disjunction. This is equivalent to saying the belief state satisfies
\begin{align*}
    & \ncode{!}\Box \code{((((y1 == 1 \&\& y\_td1 == 1) || (y2 == 1 \&\& y\_td2 == 1)) \&\& y\_radar <= 1000)}\\
                   & \code{=> altitude == 0)}
\end{align*}
While we omit the full derivation, we note that from $\existsprop^{\mathrm{MPL}}$ we can derive that the belief state satisfies the negated proposition:
\begin{align*}
    & \Box \code{((((y1 == 1 \&\& y\_td1 == 1) || (y2 == 1 \&\& y\_td2 == 1)) \&\& y\_radar <= 1000)}\\
   & \code{=> altitude == 0)}
\end{align*}
Thus, by contradiction, we can disregard this last term of the disjunction.

Because the remaining terms only contain quantified variables, we can apply Theorem~\ref{thm:envindep} to see that
\begin{align*}
    \forall \bstate . \; \bstate \vDash & \rcode{altitude != 0} \Rightarrow\\
    \bstate \vDash & \; \Box \ncode{(y\_radar > 1000) || } \\
                   & \; \Box \ncode{((y1 != 1 || y\_td1 != 1) \&\& (y2 != 1 || y\_td2 != 1))} \\
\end{align*}
We now consider each case of the disjunction separately.

In the first case, the approach is to show that for all $\bstate$,
\[
\bstate \vDash \Box \rcode{(y\_radar > 1000)} \Rightarrow
\bstate \vDash \Box \rcode{(altitude > 0)} \Rightarrow
\bstate \vDash \Box \ncode{(time\_on\_ground == 0)}
\]
where the first implication uses the assumptions on Lines \ref{code:yradar} and \ref{code:radaracc} of $\existsprop^\mathrm{MPL}$ and the second implication uses the result from Section~\ref{sec:togalt}.

In the second case, we will assume that the belief state satisfies \lstinline[breaklines=true]{$\Box$(y_radar <= 1000)}, because otherwise by Theorem~\ref{thm:envindep} we would have
\[
    \bstate \vDash \Diamond\rcode{(y\_radar > 1000)} \Rightarrow \bstate \vDash \Box\rcode{(y\_radar > 1000)}
\]
which by the above argument must ensure the result. Because of this, we can assume without reservation that \lstinline[breaklines=true]{$\Box$(landing_leg_deployment == 0)}.  
We will also assume for now that \lstinline[breaklines=true]{$\Box$(permanent_1 != 1)}. 
In any environment $\sigma$ in a belief state satisfying this, we have that
\begin{align*}
    \sigma \vDash & \code{y\_td1 != 1 \&\& transient\_on\_touchdown == 0} \Rightarrow\\
    \sigma \vDash & \code{altitude != 0} \Rightarrow \sigma \vDash \code{time\_on\_ground == 0}
\end{align*}
Where the first implication uses assumptions from Lines \ref{code:totasgn} and \ref{code:tdiasgn} and the second one uses the result from Section~\ref{sec:togalt}. Similarly, we have that
\begin{align*}
    \sigma \vDash & \code{y\_t1 != 1 \&\& transient\_on\_touchdown == 0} \Rightarrow\\
    \sigma \vDash & \code{altitude\_0 != 0} \Rightarrow%
    \sigma \vDash \code{time\_on\_ground\_0 == 0} \Rightarrow \\
    \sigma \vDash & \code{time\_on\_ground == 0}
\end{align*}
This means that in any belief state satisfying $\existsprop^\mathrm{MPL}$, we will have that
\[
    \Box\ncode{(transient\_on\_touchdown == 0 => time\_on\_ground == 0)}
\]

If the assumption \lstinline[breaklines=true]{$\Box$(permanent_1 != 1)} is not true, then by the assumption on Line~\ref{code:perm} we can assume that \lstinline[breaklines=true]{$\Box$(permanent_2 != 1)} and apply a symmetric line of reasoning to \lstinline[breaklines=true]{y_td2} and \lstinline[breaklines=true]{y2}.

\subsubsection{Bounded time-on-ground}

We will now show that 
\[
    \bstate \vDash \existsprop^\mathrm{MPL} \Rightarrow \bstate \vDash \Box 
    \ncode{(transient\_on\_touchdown => time\_on\_ground < 2)}
\]
We can summarize the proposition on Line~\ref{code:togasgn} as
\begin{align*}
    \forall \bstate. \; \bstate \vDash & \Box \ncode{%
        (engine\_enabled\_0 == 1 => time\_on\_ground == time\_on\_ground\_0 + 1 \&\&}\\
    & \rcode{engine\_enabled\_0 != 1 => time\_on\_ground == time\_on\_ground\_0)}
\end{align*}
We can now apply the fact that
\begin{align*}
    \forall \sigma. \; \sigma \vDash & \code{transient\_on\_touchdown == 0}\\
    & \code{engine\_enabled\_0 == 1 => time\_on\_ground == time\_on\_ground\_0 + 1 \&\&}\\
    & \code{engine\_enabled\_0 != 1 => time\_on\_ground == time\_on\_ground\_0 \&\&}\\
    & \code{time\_on\_ground\_0 >= 0 \&\&}\\
    & \code{time\_on\_ground\_0 < 2 \&\&}\\
    & \code{time\_on\_ground\_0 == 1 => engine\_enabled\_0 == 0}\\
    \Rightarrow \sigma \vDash & \code{transient\_on\_touchdown => time\_on\_ground < 2}
\end{align*}
can be proven using standard techniques for inequlities and propositional logic. We can lift this to the $\Box$ modality using Theorems \ref{thm:lifting} and \ref{thm:k} and then further apply the assumptions on Lines \ref{code:tognonneg}, \ref{code:togezdis}, and \ref{code:togbound} to show the result.

\section{The MPL-Exp Benchmark}
\label{sec:mplexp}

In this section, we describe the MPL-Exp benchmark.
This section is laid out similarly to Section~\ref{sec:casestudy}.
Figure~\ref{fig:mplexpmodel} presents a modified version of Figure~\ref{fig:mplmodel}, and Figure~\ref{fig:mplexpbp} presents a modified version of Figure~\ref{fig:mplbp}.
Section~\ref{sec:mplexpverification} is analogous to \ref{sec:mplverification} and explains how to verify the modified programs using Epistemic Hoare logic.

\begin{figure}
    \begin{lstlisting}[basicstyle=\ttfamily\small]
//Model Initialization
prev_err_1 = 0; prev_err_2 = 0; trans_td = 0;
alt = 4;
time_on_ground = 0;
//Permanent errors
perm_1 = choose(. == 0 || . == 1);
perm_2 = choose((. == 0 || . == 1) && (perm_1 == 1 => . == 0));
perm_1_v = choose(. == 0 || . == 1); 
perm_2_v = choose(. == 0 || . == 1);
//Controller initialization
...
while(engine_enabled == 1) 
{ (alt > -1 => engine_enabled == 1) && 
  (trans_td == 0 => time_on_ground < 2) }
{
    //Model start
    if (alt == -1 && engine_enabled == 1) {
        time_on_ground = time_on_ground + 1
    };
    if (alt >= 3) { alt_rate = choose(. == 0 || . == 1) } 
    else          { alt_rate = choose(. <= prev_alt + 1) };
    alt = alt - alt_rate;
    if (alt >= 3 ) { radar_alt = choose( alt - 1 <= . && . <= alt + 1) }
    else if (alt == 2) { radar_alt = choose ( -1 <= . && . < = 3) }
    else { radar_alt = choose(-1 <= . && . <= 2) };

    err_1 = choose((prev_err_1 == 1 => . == 0) && (. == 0 || . == 1));
    prev_err_1 = err_1;
    err_2 = choose((prev_err_2 == 1 => . == 0) && (. == 0 || . == 1));
    prev_err_2 = err_2;
    if (alt == -1 && (err_1 == 1 || err_2 == 1)) { trans_td = 1; };

    leg_err = alt == 4;
    if perm_1 { cur_td_1 = perm_1_v } 
    else if (leg_err == 1 || err_1 == 1) { 
        cur_td_1 = choose(. == 0 || . == 1)
    } else { cur_td_1 = alt == -1; }
    else if (leg_err == 1 || err_2 == 1)  {
        cur_td_2 = choose(. == 0 || . == 1)
    } else { cur_td_2 = alt == -1; };

    //Model end
    //Controller loop start
    ...
}
    \end{lstlisting}
    \caption{Model for the MPL-Exp benchmark.}
    \label{fig:mplexpmodel}
\end{figure}

\begin{figure}
    \begin{lstlisting}[basicstyle=\ttfamily\small]
// Model Initialization
...
engine_enabled = 1;
while (engine_enabled == 1)
{ $\Box$((alt > -1 => engine_enabled == 1) && 
    (trans_td == 0 => time_on_ground < 2)) }
{
    // Model start
    ...
    // Model end
    observe radar_alt;
    observe cur_td_1;
    observe cur_td_2;

    infer $\Box$(alt == -1) {
        engine_enabled = 0
    }
}
    \end{lstlisting}
    \caption{Belief program of the MPL-Exp benchmark.}
    \label{fig:mplexpbp}
\end{figure}

\paragraph{\bf Altitude Model.} Instead of the uniform 1-meter discretization in Figure~\ref{fig:mplmodel}, the MPL-Exp benchmark uses an exponential discretization. This modified discretization applies to the \texttt{alt} and \texttt{radar\_alt} variables, and is defined as follows:
\begin{itemize}
    \item When an altitude variable is 4, the altitude is between 1000 and 10000 meters.
    \item When an altitude variable is 3, the altitude is between 100 and 1000 meters.
    \item When an altitude variable is 2, the altitude is between 10 and 100 meters.
    \item When an altitude variable is 1, the altitude is between 1 and 10 meters.
    \item When an altitude variable is 0, the altitude is between 0 and 1 meters.
    \item When an altitude variable is -1, the altitude is exactly 0 meters.
\end{itemize}
The model in Figure~\ref{fig:mplexpmodel} is the same as that of Figure~\ref{fig:mplmodel} except that the altitude model has been modified to reflect the exponential discretization.
The update rules are designed to conservatively over-approximate the model in Figure~\ref{fig:mplexpmodel}.
This means that after applying the definition of the discretization, any possible true environment in Figure~\ref{fig:mplmodel} is also possible under Figure~\ref{fig:mplexpmodel}.

The belief program in Figure~\ref{fig:mplexpbp} is the same as that of Figure~\ref{fig:mplbp} except that all references to the \texttt{alt} variable have been modified to use the exponential discretization.

\subsection{Verification}
\label{sec:mplexpverification}

In this section, we explain how to verify the loop invariant of the MPL-Exp example, which describes the same property as that of the MPL example, but with the exponential altitude discretization.
The process proceeds the same as in Section~\ref{sec:mplverification}. We first construct a strengthened invariant, which is the same as the strengthened invariant in Section~\ref{sec:mplverification} but with altitude comparisons to \texttt{0} replaced with comparisons to \texttt{-1}.
Then, using the rules of Epistemic Hoare logic, we construct a formula $\existsprop^\mathrm{MPL-Exp}$ by propagating the strengthened invariant through the loop body, and show that $\existsprop^\mathrm{MPL-Exp}$ implies the invariant itself.

To show that $\existsprop^\mathrm{MPL-Exp}$ implies the strengthened invariant, we follow the same general line of reasoning as in Section~\ref{sec:mplverification}. 
All of the proof except the engine disabled soundness condition follows the exact same reasoning as Section~\ref{sec:mplverification} with appropriate changes to take into account the new strengthened invariant and the use of $\existsprop^\mathrm{MPL-Exp}$.

Showing engine disabled soundness requires more care because the proof makes use of intermediate propositions that critically depend on the altitude model.

\subsubsection{Engine Disabled Soundness}

To apply an analogous line of reasoning to Section~\ref{sec:engdissound}, we need a value $\num^*$ that satisfies the conditions 
$
\bstate \vDash \Box \code{(y\_radar >} \num^* \rcode{)} \Rightarrow
\bstate \vDash \Box \rcode{(alt > -1)}
$
and
\lstinline[breaklines=true]{$\bstate \vDash \Box$(y_radar <= $\num^*$) $\Rightarrow$ $\bstate \vDash \Box$(landing_leg_deployment == 0)}.

While we omit the full derivation here, we note that choosing $\num^* = \texttt{2}$, applying $\existsprop^\mathrm{MPL-Exp}$ and Theorems \ref{thm:lifting} and \ref{thm:k} can prove these implications.

The remaining pieces of the proof are the same as in Section~\ref{sec:engdissound}.

\end{document}